%% file: 20OJCAS_BOX1OFDM_main.tex
\documentclass[journal,twoside,twocolumn]{IEEEtran}
\usepackage{cite}
\usepackage[T1]{fontenc}
\usepackage{graphicx}
\usepackage{amssymb}
\usepackage{amsmath}
\usepackage{paralist}
\usepackage{subfigure}
\usepackage{booktabs} 
\usepackage{algorithm}
\usepackage{amsthm}
\usepackage{multirow}
\usepackage{microtype} 
\usepackage{tablefootnote}
\usepackage{scrextend}
\usepackage{balance}
\usepackage{algpseudocode}
\usepackage{hyperref}

\input{vmr-symbols-vecbold}
\input{standard-macros}

\input{defs}

\usepackage{color}

\usepackage{enumitem}
\setlist[itemize]{leftmargin=*, itemsep=0.3em, topsep=0.3em} 

\allowdisplaybreaks

\newcommand{\revision}[1]{{#1}}
\newcommand{\revisionS}[1]{{#1}}

\safemath{\LAMA}{\textrm{LAMA}}
\safemath{\MRT}{\textrm{MRT}}
\safemath{\betamax}{\beta^\text{max}_\setO}
\safemath{\betamaxno}{\beta^\text{max}}
\safemath{\betamin}{\beta^\text{min}_\setO}
\safemath{\betaminno}{\beta^\text{min}}

\safemath{\Nomin}{\No^\textnormal{min}(\beta)}
\safemath{\Nominnobeta}{\No^\text{min}}
\safemath{\Nomax}{\No^\textnormal{max}(\beta)}
\safemath{\Nomaxnobeta}{\No^\textnormal{max}}
\safemath{\EX}{E_\textnormal{x}}
\safemath{\EXP}{\EX^\textnormal{p}}
\safemath{\Eo}{E_0}

\safemath{\tmax}{{t_\textnormal{max}}}
\safemath{\MAP}{\textrm{MAP}}
\safemath{\IO}{\textrm{IO}}
\safemath{\JO}{\textrm{JO}}
\safemath{\Nopost}{N_{0}^\textnormal{post}}
\safemath{\MT}{U}
\safemath{\MR}{B}
\safemath{\Tran}{\textnormal{T}}
\safemath{\Herm}{\textnormal{H}}
\safemath{\row}{\textnormal{r}}
\safemath{\col}{\textnormal{c}}

\safemath{\NT}{N_\textnormal{T}}
\safemath{\DSNR}{\delta \textnormal{SNR}}
\safemath{\betaMOR}{\beta^{\star}}
\newcommand{\mymath}[1]{\mathrm{#1}}
\newcommand{\mymathb}[1]{\boldsymbol{\mathrm{#1}}}

\DeclareMathOperator{\diag}{diag}
\DeclareMathOperator{\proj}{proj}
\newtheorem{remark}{Remark}

\begin{document}

\title{Algorithm and VLSI Design for 1-bit Data Detection in  Massive MIMO-OFDM}
\author{Seyed Hadi Mirfarshbafan, Mahdi Shabany, Seyed Alireza Nezamalhosseini, and Christoph Studer
\thanks{Parts of this paper have been presented at IEEE ISCAS 2018 \cite{mirfarshbafan18};
 in this paper, we extend our results to frequency-selective channels using OFDM.}
\thanks{\revision{S.~H.~Mirfarshbafan was with the School of Electrical and Computer Engineering at Cornell Tech, New York, NY.}}
\thanks{M.~Shabany is with the Department of Electrical Engineering, Sharif University of Technology, Tehran, Iran; S.~A.~Nezamalhosseini is with the Department of Electrical Engineering, Iran University of Science and Technology, Tehran, Iran.}
\thanks{\revision{C.~Studer was with the School of Electrical and Computer Engineering at Cornell Tech, New York, NY, and Cornell University, Ithaca, NY, and is now with the Department of Information Technology and Electrical Engineering at ETH Zurich, Switzerland. Email: studer@ethz.ch; Web: \url{http://iis.ee.ethz.ch}}}
\thanks{The work of SHM and CS was supported in part by Xilinx, Inc., and the US National Science Foundation (NSF) under grants ECCS-1408006, CCF-1535897,  CCF-1652065, CNS-1717559, and ECCS-1824379.}
\thanks{A MATLAB simulator to reproduce the results of this paper is
available on GitHub: \url{https://github.com/IIP-Group/1BOX-MIMO-OFDM-simulator}}
}

\maketitle
\input{0-abstract}

\input{1-introduction}
\input{2-system_model}

\input{3-data_detection_channel_estimation}
\input{4-VLSI}

\input{5-conclusion}

\balance
\bibliographystyle{IEEEtran}
\bibliography{bib/confs-jrnls,bib/IEEEabrv,bib/publishers,bib/vipbib}
\balance

\end{document}

%% file: vmr-symbols-vecbold.tex
\usepackage{amssymb}
\usepackage{amsfonts}
\usepackage{mathrsfs}
\usepackage{xspace}
\usepackage{bm}
\usepackage{upgreek}

\newcommand{\safemath}[2]{\newcommand{#1}{\ensuremath{#2}\xspace}}

\safemath{\bma}{\mathbf{a}}
\safemath{\bmb}{\mathbf{b}}
\safemath{\bmc}{\mathbf{c}}
\safemath{\bmd}{\mathbf{d}}
\safemath{\bme}{\mathbf{e}}
\safemath{\bmf}{\mathbf{f}}
\safemath{\bmg}{\mathbf{g}}
\safemath{\bmh}{\mathbf{h}}
\safemath{\bmi}{\mathbf{i}}
\safemath{\bmj}{\mathbf{j}}
\safemath{\bmk}{\mathbf{k}}
\safemath{\bml}{\mathbf{l}}
\safemath{\bmm}{\mathbf{m}}
\safemath{\bmn}{\mathbf{n}}
\safemath{\bmo}{\mathbf{o}}
\safemath{\bmp}{\mathbf{p}}
\safemath{\bmq}{\mathbf{q}}
\safemath{\bmr}{\mathbf{r}}
\safemath{\bms}{\mathbf{s}}
\safemath{\bmt}{\mathbf{t}}
\safemath{\bmu}{\mathbf{u}}
\safemath{\bmv}{\mathbf{v}}
\safemath{\bmw}{\mathbf{w}}
\safemath{\bmx}{\mathbf{x}}
\safemath{\bmy}{\mathbf{y}}
\safemath{\bmz}{\mathbf{z}}
\safemath{\bmzero}{\mathbf{0}}
\safemath{\bmone}{\mathbf{1}}

\bmdefine{\biad}{a}
\bmdefine{\bibd}{b}
\bmdefine{\bicd}{c}
\bmdefine{\bidd}{d}
\bmdefine{\bied}{e}
\bmdefine{\bifd}{f}
\bmdefine{\bigd}{g}
\bmdefine{\bihd}{h}
\bmdefine{\biid}{i}
\bmdefine{\bijd}{j}
\bmdefine{\bikd}{k}
\bmdefine{\bild}{l}
\bmdefine{\bimd}{m}
\bmdefine{\bind}{n}
\bmdefine{\biod}{o}
\bmdefine{\bipd}{p}
\bmdefine{\biqd}{q}
\bmdefine{\bird}{r}
\bmdefine{\bisd}{s}
\bmdefine{\bitd}{t}
\bmdefine{\biud}{u}
\bmdefine{\bivd}{v}
\bmdefine{\biwd}{w}
\bmdefine{\bixd}{x}
\bmdefine{\biyd}{y}
\bmdefine{\bizd}{z}

\bmdefine{\bixid}{\xi}
\bmdefine{\bilambdad}{\lambda}
\bmdefine{\bimud}{\mu}
\bmdefine{\bithetad}{\theta}
\bmdefine{\biphid}{\phi}
\bmdefine{\bideltad}{\delta}

\safemath{\bmia}{\biad}
\safemath{\bmib}{\bibd}
\safemath{\bmic}{\bicd}
\safemath{\bmid}{\bidd}
\safemath{\bmie}{\bied}
\safemath{\bmif}{\bifd}
\safemath{\bmig}{\bigd}
\safemath{\bmih}{\bihd}
\safemath{\bmii}{\biid}
\safemath{\bmij}{\bijd}
\safemath{\bmik}{\bikd}
\safemath{\bmil}{\bild}
\safemath{\bmim}{\bimd}
\safemath{\bmin}{\bind}
\safemath{\bmio}{\biod}
\safemath{\bmip}{\bipd}
\safemath{\bmiq}{\biqd}
\safemath{\bmir}{\bird}
\safemath{\bmis}{\bisd}
\safemath{\bmit}{\bitd}
\safemath{\bmiu}{\biud}
\safemath{\bmiv}{\bivd}
\safemath{\bmiw}{\biwd}
\safemath{\bmix}{\bixd}
\safemath{\bmiy}{\biyd}
\safemath{\bmiz}{\bizd}

\safemath{\bmxi}{\bixid}
\safemath{\bmlambda}{\bilambdad}
\safemath{\bmmu}{\bimud}
\safemath{\bmtheta}{\bithetad}
\safemath{\bmphi}{\biphid}
\safemath{\bmdelta}{\bideltad}

\safemath{\bA}{\mathbf{A}}
\safemath{\bB}{\mathbf{B}}
\safemath{\bC}{\mathbf{C}}
\safemath{\bD}{\mathbf{D}}
\safemath{\bE}{\mathbf{E}}
\safemath{\bF}{\mathbf{F}}
\safemath{\bG}{\mathbf{G}}
\safemath{\bH}{\mathbf{H}}
\safemath{\bI}{\mathbf{I}}
\safemath{\bJ}{\mathbf{J}}
\safemath{\bK}{\mathbf{K}}
\safemath{\bL}{\mathbf{L}}
\safemath{\bM}{\mathbf{M}}
\safemath{\bN}{\mathbf{N}}
\safemath{\bO}{\mathbf{O}}
\safemath{\bP}{\mathbf{P}}
\safemath{\bQ}{\mathbf{Q}}
\safemath{\bR}{\mathbf{R}}
\safemath{\bS}{\mathbf{S}}
\safemath{\bT}{\mathbf{T}}
\safemath{\bU}{\mathbf{U}}
\safemath{\bV}{\mathbf{V}}
\safemath{\bW}{\mathbf{W}}
\safemath{\bX}{\mathbf{X}}
\safemath{\bY}{\mathbf{Y}}
\safemath{\bZ}{\mathbf{Z}}

\safemath{\bZero}{\mathbf{0}}
\safemath{\bOne}{\mathbf{1}}
\safemath{\bDelta}{\mathbf{\Delta}}
\safemath{\bLambda}{\mathbf{\UpLambda}}
\safemath{\bPhi}{\mathbf{\Upphi}}
\safemath{\bSigma}{\mathbf{\Upsigma}}
\safemath{\bOmega}{\mathbf{\Upomega}}
\safemath{\bTheta}{\mathbf{\Uptheta}}

\bmdefine{\biAd}{A}
\bmdefine{\biBd}{B}
\bmdefine{\biCd}{C}
\bmdefine{\biDd}{D}
\bmdefine{\biEd}{E}
\bmdefine{\biFd}{F}
\bmdefine{\biGd}{G}
\bmdefine{\biHd}{H}
\bmdefine{\biId}{I}
\bmdefine{\biJd}{J}
\bmdefine{\biKd}{K}
\bmdefine{\biLd}{L}
\bmdefine{\biMd}{M}
\bmdefine{\biOd}{N}
\bmdefine{\biPd}{O}
\bmdefine{\biQd}{P}
\bmdefine{\biRd}{R}
\bmdefine{\biSd}{S}
\bmdefine{\biTd}{T}
\bmdefine{\biUd}{U}
\bmdefine{\biVd}{V}
\bmdefine{\biWd}{W}
\bmdefine{\biXd}{X}
\bmdefine{\biYd}{Y}
\bmdefine{\biZd}{Z}

\bmdefine{\biDelta}{\Delta}
\bmdefine{\biLambda}{\Lambda}
\bmdefine{\biPhi}{\Phi}
\bmdefine{\biSigma}{\Sigma}
\bmdefine{\biOmega}{\Omega}
\bmdefine{\biTheta}{\Theta}

\safemath{\bimA}{\biAd}
\safemath{\bimB}{\biBd}
\safemath{\bimC}{\biCd}
\safemath{\bimD}{\biDd}
\safemath{\bimE}{\biEd}
\safemath{\bimF}{\biFd}
\safemath{\bimG}{\biGd}
\safemath{\bimH}{\biHd}
\safemath{\bimI}{\biId}
\safemath{\bimJ}{\biJd}
\safemath{\bimK}{\biKd}
\safemath{\bimL}{\biLd}
\safemath{\bimM}{\biMd}
\safemath{\bimN}{\biNd}
\safemath{\bimO}{\biOd}
\safemath{\bimP}{\biPd}
\safemath{\bimQ}{\biQd}
\safemath{\bimR}{\biRd}
\safemath{\bimS}{\biSd}
\safemath{\bimT}{\biTd}
\safemath{\bimU}{\biUd}
\safemath{\bimV}{\biVd}
\safemath{\bimW}{\biWd}
\safemath{\bimX}{\biXd}
\safemath{\bimY}{\biYd}
\safemath{\bimZ}{\biZd}

\safemath{\bimDelta}{\biDelta}
\safemath{\bimLambda}{\biLambda}
\safemath{\bimPhi}{\biPhi}
\safemath{\bimSigma}{\biSigma}
\safemath{\bimOmega}{\biOmega}
\safemath{\bimTheta}{\biTheta}

\safemath{\setA}{\mathcal{A}}
\safemath{\setB}{\mathcal{B}}
\safemath{\setC}{\mathcal{C}}
\safemath{\setD}{\mathcal{D}}
\safemath{\setE}{\mathcal{E}}
\safemath{\setF}{\mathcal{F}}
\safemath{\setG}{\mathcal{G}}
\safemath{\setH}{\mathcal{H}}
\safemath{\setI}{\mathcal{I}}
\safemath{\setJ}{\mathcal{J}}
\safemath{\setK}{\mathcal{K}}
\safemath{\setL}{\mathcal{L}}
\safemath{\setM}{\mathcal{M}}
\safemath{\setN}{\mathcal{N}}
\safemath{\setO}{\mathcal{O}}
\safemath{\setP}{\mathcal{P}}
\safemath{\setQ}{\mathcal{Q}}
\safemath{\setR}{\mathcal{R}}
\safemath{\setS}{\mathcal{S}}
\safemath{\setT}{\mathcal{T}}
\safemath{\setU}{\mathcal{U}}
\safemath{\setV}{\mathcal{V}}
\safemath{\setW}{\mathcal{W}}
\safemath{\setX}{\mathcal{X}}
\safemath{\setY}{\mathcal{Y}}
\safemath{\setZ}{\mathcal{Z}}
\safemath{\emptySet}{\varnothing}

\safemath{\colA}{\mathscr{A}}
\safemath{\colB}{\mathscr{B}}
\safemath{\colC}{\mathscr{C}}
\safemath{\colD}{\mathscr{D}}
\safemath{\colE}{\mathscr{E}}
\safemath{\colF}{\mathscr{F}}
\safemath{\colG}{\mathscr{G}}
\safemath{\colH}{\mathscr{H}}
\safemath{\colI}{\mathscr{I}}
\safemath{\colJ}{\mathscr{J}}
\safemath{\colK}{\mathscr{K}}
\safemath{\colL}{\mathscr{L}}
\safemath{\colM}{\mathscr{M}}
\safemath{\colN}{\mathscr{N}}
\safemath{\colO}{\mathscr{O}}
\safemath{\colP}{\mathscr{P}}
\safemath{\colQ}{\mathscr{Q}}
\safemath{\colR}{\mathscr{R}}
\safemath{\colS}{\mathscr{S}}
\safemath{\colT}{\mathscr{T}}
\safemath{\colU}{\mathscr{U}}
\safemath{\colV}{\mathscr{V}}
\safemath{\colW}{\mathscr{W}}
\safemath{\colX}{\mathscr{X}}
\safemath{\colY}{\mathscr{Y}}
\safemath{\colZ}{\mathscr{Z}}

\safemath{\opA}{\mathbb{A}}
\safemath{\opB}{\mathbb{B}}
\safemath{\opC}{\mathbb{C}}
\safemath{\opD}{\mathbb{D}}
\safemath{\opE}{\mathbb{E}}
\safemath{\opF}{\mathbb{F}}
\safemath{\opG}{\mathbb{G}}
\safemath{\opH}{\mathbb{H}}
\safemath{\opI}{\mathbb{I}}
\safemath{\opJ}{\mathbb{J}}
\safemath{\opK}{\mathbb{K}}
\safemath{\opL}{\mathbb{L}}
\safemath{\opM}{\mathbb{M}}
\safemath{\opN}{\mathbb{N}}
\safemath{\opO}{\mathbb{O}}
\safemath{\opP}{\mathbb{P}}
\safemath{\opQ}{\mathbb{Q}}
\safemath{\opR}{\mathbb{R}}
\safemath{\opS}{\mathbb{S}}
\safemath{\opT}{\mathbb{T}}
\safemath{\opU}{\mathbb{U}}
\safemath{\opV}{\mathbb{V}}
\safemath{\opW}{\mathbb{W}}
\safemath{\opX}{\mathbb{X}}
\safemath{\opY}{\mathbb{Y}}
\safemath{\opZ}{\mathbb{Z}}
\safemath{\opZero}{\mathbb{O}}
\safemath{\identityop}{\opI}

\safemath{\veca}{\bma}
\safemath{\vecb}{\bmb}
\safemath{\vecc}{\bmc}
\safemath{\vecd}{\bmd}
\safemath{\vece}{\bme}
\safemath{\vecf}{\bmf}
\safemath{\vecg}{\bmg}
\safemath{\vech}{\bmh}
\safemath{\veci}{\bmi}
\safemath{\vecj}{\bmj}
\safemath{\veck}{\bmk}
\safemath{\vecl}{\bml}
\safemath{\vecm}{\bmm}
\safemath{\vecn}{\bmn}
\safemath{\veco}{\bmo}
\safemath{\vecp}{\bmp}
\safemath{\vecq}{\bmq}
\safemath{\vecr}{\bmr}
\safemath{\vecs}{\bms}
\safemath{\vect}{\bmt}
\safemath{\vecu}{\bmu}
\safemath{\vecv}{\bmv}
\safemath{\vecw}{\bmw}
\safemath{\vecx}{\bmx}
\safemath{\vecy}{\bmy}
\safemath{\vecz}{\bmz}

\safemath{\veczero}{\bmzero}
\safemath{\vecone}{\bmone}
\safemath{\vecxi}{\bmxi}
\safemath{\veclambda}{\bmlambda}
\safemath{\vecmu}{\bmmu}
\safemath{\vectheta}{\bmtheta}
\safemath{\vecphi}{\bmphi}
\safemath{\vecdelta}{\bmdelta}

\safemath{\matA}{\bA}
\safemath{\matB}{\bB}
\safemath{\matC}{\bC}
\safemath{\matD}{\bD}
\safemath{\matE}{\bE}
\safemath{\matF}{\bF}
\safemath{\matG}{\bG}
\safemath{\matH}{\bH}
\safemath{\matI}{\bI}
\safemath{\matJ}{\bJ}
\safemath{\matK}{\bK}
\safemath{\matL}{\bL}
\safemath{\matM}{\bM}
\safemath{\matN}{\bN}
\safemath{\matO}{\bO}
\safemath{\matP}{\bP}
\safemath{\matQ}{\bQ}
\safemath{\matR}{\bR}
\safemath{\matS}{\bS}
\safemath{\matT}{\bT}
\safemath{\matU}{\bU}
\safemath{\matV}{\bV}
\safemath{\matW}{\bW}
\safemath{\matX}{\bX}
\safemath{\matY}{\bY}
\safemath{\matZ}{\bZ}
\safemath{\matzero}{\bmzero}

\safemath{\matDelta}{\bDelta}
\safemath{\matLambda}{\bLambda}
\safemath{\matPhi}{\bPhi}
\safemath{\matSigma}{\bSigma}
\safemath{\matOmega}{\bOmega}
\safemath{\matTheta}{\bTheta}

\safemath{\matidentity}{\matI}
\safemath{\matone}{\matO}

\safemath{\rnda}{A}
\safemath{\rndb}{B}
\safemath{\rndc}{C}
\safemath{\rndd}{D}
\safemath{\rnde}{E}
\safemath{\rndf}{F}
\safemath{\rndg}{G}
\safemath{\rndh}{H}
\safemath{\rndi}{I}
\safemath{\rndj}{J}
\safemath{\rndk}{K}
\safemath{\rndl}{L}
\safemath{\rndm}{M}
\safemath{\rndn}{N}
\safemath{\rndo}{O}
\safemath{\rndp}{P}
\safemath{\rndq}{Q}
\safemath{\rndr}{R}
\safemath{\rnds}{S}
\safemath{\rndt}{T}
\safemath{\rndu}{U}
\safemath{\rndv}{V}
\safemath{\rndw}{W}
\safemath{\rndx}{X}
\safemath{\rndy}{Y}
\safemath{\rndz}{Z}

\safemath{\rveca}{\bimA}
\safemath{\rvecb}{\bimB}
\safemath{\rvecc}{\bimC}
\safemath{\rvecd}{\bimD}
\safemath{\rvece}{\bimE}
\safemath{\rvecf}{\bimF}
\safemath{\rvecg}{\bimG}
\safemath{\rvech}{\bimH}
\safemath{\rveci}{\bimI}
\safemath{\rvecj}{\bimJ}
\safemath{\rveck}{\bimK}
\safemath{\rvecl}{\bimL}
\safemath{\rvecm}{\bimM}
\safemath{\rvecn}{\bimN}
\safemath{\rveco}{\bomO}
\safemath{\rvecp}{\bimP}
\safemath{\rvecq}{\bimQ}
\safemath{\rvecr}{\bimR}
\safemath{\rvecs}{\bimS}
\safemath{\rvect}{\bimT}
\safemath{\rvecu}{\bimU}
\safemath{\rvecv}{\bimV}
\safemath{\rvecw}{\bimW}
\safemath{\rvecx}{\bimX}
\safemath{\rvecy}{\bimY}
\safemath{\rvecz}{\bimZ}

\safemath{\rvecxi}{\bmxi}
\safemath{\rveclambda}{\bmlambda}
\safemath{\rvecmu}{\bmmu}
\safemath{\rvectheta}{\bmtheta}
\safemath{\rvecphi}{\bmphi}

\safemath{\rmatA}{\bimA}
\safemath{\rmatB}{\bimB}
\safemath{\rmatC}{\bimC}
\safemath{\rmatD}{\bimD}
\safemath{\rmatE}{\bimE}
\safemath{\rmatF}{\bimF}
\safemath{\rmatG}{\bimG}
\safemath{\rmatH}{\bimH}
\safemath{\rmatI}{\bimI}
\safemath{\rmatJ}{\bimJ}
\safemath{\rmatK}{\bimK}
\safemath{\rmatL}{\bimL}
\safemath{\rmatM}{\bimM}
\safemath{\rmatN}{\bimN}
\safemath{\rmatO}{\bimO}
\safemath{\rmatP}{\bimP}
\safemath{\rmatQ}{\bimQ}
\safemath{\rmatR}{\bimR}
\safemath{\rmatS}{\bimS}
\safemath{\rmatT}{\bimT}
\safemath{\rmatU}{\bimU}
\safemath{\rmatV}{\bimV}
\safemath{\rmatW}{\bimW}
\safemath{\rmatX}{\bimX}
\safemath{\rmatY}{\bimY}
\safemath{\rmatZ}{\bimZ}

\safemath{\rmatDelta}{\bimDelta}
\safemath{\rmatLambda}{\bimLambda}
\safemath{\rmatPhi}{\bimPhi}
\safemath{\rmatSigma}{\bimSigma}
\safemath{\rmatOmega}{\bimOmega}
\safemath{\rmatTheta}{\bimTheta}

%% file: standard-macros.tex
\usepackage{amssymb}
\usepackage{amsfonts}
\usepackage{mathrsfs}
\usepackage{xspace}
\usepackage{bm}
\usepackage{fancyref}
\usepackage{textcomp}

\usepackage{multirow}
\usepackage{stmaryrd}

\newenvironment{textbmatrix}{	\setlength{\arraycolsep}{2.5pt}%
								\big[\begin{matrix}}{\end{matrix}\big]%
								\raisebox{0.08ex}{\vphantom{M}}}

\def\be{\begin{equation}}
\def\ee{\end{equation}}
\def\een{\nonumber \end{equation}}
\def\mat{\begin{bmatrix}}
\def\emat{\end{bmatrix}}
\def\btm{\begin{textbmatrix}}
\def\etm{\end{textbmatrix}}

\def\ba#1\ea{\begin{align}#1\end{align}}
\def\bas#1\eas{\begin{align*}#1\end{align*}}
\def\bs#1\es{\begin{split}#1\end{split}}
\def\bg#1\eg{\begin{gather}#1\end{gather}}
\def\bml#1\eml{\begin{multline}#1\end{multline}}
\def\bi#1\ei{\begin{itemize}#1\end{itemize}}

\newcommand{\lefto}{\mathopen{}\left}

\DeclareMathOperator{\sign}{sign}			
\DeclareMathOperator*{\argmin}{arg\;min}		
\DeclareMathOperator*{\argmax}{arg\;max}		
\DeclareMathOperator{\Exop}{\opE}			

\newcommand{\Ex}[2]{\ensuremath{\Exop_{#1}\lefto[#2\right]}} 	

\safemath{\dirac}{\delta}					
\safemath{\krond}{\dirac}					

\safemath{\upto}{\uparrow}
\safemath{\downto}{\downarrow}
\safemath{\iu}{j}							
\safemath{\ev}{\lambda}						
\safemath{\hilseqspace}{l^{2}}				
\newcommand{\banachfunspace}[1]{\setL^{#1}}	
\safemath{\hilfunspace}{\banachfunspace{2}}	

\safemath{\SNR}{\textit{SNR}} 				
\safemath{\PAR}{\textit{PAR}} 				
\safemath{\No}{N_0}							
\safemath{\Es}{E_s}							
\safemath{\Eb}{E_b}							
\safemath{\EbNo}{\frac{\Eb}{\No}}
\safemath{\EsNo}{\frac{\Es}{\No}}

\DeclareMathOperator{\CHop}{\ensuremath{\opH}} 
\safemath{\tvir}{\rndh_{\CHop}}				
\safemath{\tvtf}{\rndl_{\CHop}}				
\safemath{\spf}{\rnds_{\CHop}}				
\safemath{\bff}{H_{\CHop}}					

\safemath{\ircf}{r_{h}}						
\safemath{\tftvcf}{r_{s}}					
\safemath{\tfcf}{r_{l}}						
\safemath{\bfcf}{r_{H}}						

\safemath{\tcorr}{c_h}						
\safemath{\scf}{c_{s}}						
\safemath{\tfcorr}{c_{l}}					
\safemath{\fcorr}{c_{H}}						

\safemath{\mi}{I}							
\safemath{\capacity}{C}						

\safemath{\normal}{\mathcal{N}}			
\safemath{\jpg}{\mathcal{CN}}			
\safemath{\mchain}{\leftrightarrow}		

\safemath{\dB}{\,\mathrm{dB}}
\safemath{\dBm}{\,\mathrm{dBm}}
\safemath{\Hz}{\,\mathrm{Hz}}
\safemath{\kHz}{\,\mathrm{kHz}}
\safemath{\MHz}{\,\mathrm{MHz}}
\safemath{\GHz}{\,\mathrm{GHz}}
\safemath{\s}{\,\mathrm{s}}
\safemath{\ms}{\,\mathrm{ms}}
\safemath{\mus}{\,\mathrm{\text{\textmu}s}}
\safemath{\ns}{\,\mathrm{ns}}
\safemath{\ps}{\,\mathrm{ps}}
\safemath{\meter}{\,\mathrm{m}}
\safemath{\mm}{\,\mathrm{mm}}
\safemath{\cm}{\,\mathrm{cm}}
\safemath{\m}{\,\mathrm{m}}
\safemath{\W}{\,\mathrm{W}}
\safemath{\mW}{\, \mathrm{mW}}
\safemath{\J}{\,\mathrm{J}}
\safemath{\K}{\,\mathrm{K}}
\safemath{\bit}{\,\mathrm{bit}}
\safemath{\nat}{\,\mathrm{nat}}

\safemath{\define}{\triangleq}			

\safemath{\equivalent}{\sim}
\safemath{\distas}{\sim}					
\safemath{\sdiff}{\Delta}				

\safemath{\reals}{\mathbb{R}}
\safemath{\positivereals}{\reals_{+}}
\safemath{\integers}{\mathbb{Z}}
\safemath{\posint}{\integers_{+}}
\safemath{\naturals}{\mathbb{N}}
\safemath{\posnaturals}{\naturals_{+}}
\safemath{\complexset}{\mathbb{C}}
\safemath{\rationals}{\mathbb{Q}}

\newcommand*{\fancyrefapplabelprefix}{app}		
\newcommand*{\fancyrefthmlabelprefix}{thm}		
\newcommand*{\fancyreflemlabelprefix}{lem}		
\newcommand*{\fancyrefcorlabelprefix}{cor}		
\newcommand*{\fancyrefdeflabelprefix}{def}		
\newcommand*{\fancyrefproplabelprefix}{prop}		
\newcommand*{\fancyrefexmpllabelprefix}{exmpl}
\newcommand*{\fancyrefalglabelprefix}{alg}		
\newcommand*{\fancyreftbllabelprefix}{tbl}		

\frefformat{vario}{\fancyrefseclabelprefix}{Section~#1}
\frefformat{vario}{\fancyrefthmlabelprefix}{Theorem~#1}
\frefformat{vario}{\fancyreftbllabelprefix}{Table~#1}
\frefformat{vario}{\fancyreflemlabelprefix}{Lemma~#1}
\frefformat{vario}{\fancyrefcorlabelprefix}{Corollary~#1}
\frefformat{vario}{\fancyrefdeflabelprefix}{Definition~#1}
\frefformat{vario}{\fancyreffiglabelprefix}{Figure~#1}
\frefformat{vario}{\fancyrefapplabelprefix}{Appendix~#1}
\frefformat{vario}{\fancyrefeqlabelprefix}{(#1)}
\frefformat{vario}{\fancyrefproplabelprefix}{Proposition~#1}
\frefformat{vario}{\fancyrefexmpllabelprefix}{Example~#1}
\frefformat{vario}{\fancyrefalglabelprefix}{Algorithm~#1}

%% file: defs.tex
\safemath{\dictab}{[\,\dicta\,\,\dictb\,]}

\safemath{\ysig}{\bmy}
\safemath{\ysighat}{\hat{\ysig}}
\safemath{\ysigdim}{M}
\safemath{\xsig}{\bmx}
\safemath{\xsigdim}{N}
\safemath{\nx}{n_x}
\safemath{\zsig}{\bmz}
\safemath{\zsigdim}{\ysigdim}
\safemath{\rsig}{\bmr}
\safemath{\Adict}{\bA}
\safemath{\Adicttilde}{\widetilde{\Adict}}
\safemath{\Adictdim}{\outputdim\times\xsigdim}
\safemath{\avec}{\bma}
\safemath{\avectilde}{\tilde{\avec}}
\safemath{\Bdict}{\bB}
\safemath{\Bdicttilde}{\widetilde{\Bdict}}
\safemath{\Cdict}{\bC}
\safemath{\cvec}{\bmc}
\safemath{\Ddict}{\bD}
\safemath{\Ddictdim}{\ysigdim\times\xsigdim}
\safemath{\dvec}{\bmd}
\safemath{\Ddicttilde}{\widetilde{\bD}}
\safemath{\Bonb}{\bB}
\safemath{\bvec}{\bmb}
\safemath{\Bonbdim}{\ysigdim\times\ysigdim}
\safemath{\noise}{\bmn}
\safemath{\noisedim}{\ysigim}
\safemath{\err}{\bme}
\safemath{\errdim}{\ysigdim}
\safemath{\errset}{\setE}
\safemath{\nerr}{n_e}
\safemath{\delop}{\bP_\errset}
\safemath{\delopc}{\bP_{{\errset}^c}}

\safemath{\cplxi}{\imath}
\safemath{\cplxj}{\jmath}

\safemath{\dict}{\matD}
\safemath{\inputdim}{N}		
\safemath{\outputdim}{M}		
\safemath{\sparsity}{S}	
\safemath{\inputdimA}{{N_a}}	
\safemath{\inputdimB}{{N_b}}	
\safemath{\elemA}{{n_a}}	
\safemath{\elemB}{{n_b}}	
\safemath{\resA}{\matR_a}	
\safemath{\resB}{\matR_b}	
\safemath{\subD}{\matS} 
\safemath{\subA}{\matS_a} 
\safemath{\subB}{\matS_b} 
\safemath{\dicta}{\matA} 	
\safemath{\dictb}{\matB} 	
\safemath{\hollowS}{H}
\safemath{\hollowA}{H_a}
\safemath{\hollowB}{H_b}
\safemath{\cross}{Z}
\safemath{\coh}{\mu_d}			
\safemath{\coha}{\mu_a}			
\safemath{\cohb}{\mu_b}			
\safemath{\mubs}{\nu}	
\safemath{\cohm}{\mu_m} 
\safemath{\dictset}{\setD}	
\safemath{\dictsetp}{\dictset(\coh,\coha,\cohb)}	
\safemath{\dictsetgen}{\dictset_\text{gen}}
\safemath{\dictsetgenp}{\dictsetgen(\coh)}
\safemath{\dictsetonb}{\dictset_\text{onb}}
\safemath{\dictsetonbp}{\dictsetonb(\coh)}

\safemath{\leftside}{U}
\safemath{\rightsideA}{R_a}
\safemath{\rightsideB}{R_b}

\safemath{\indexS}{\setI_S} 

\safemath{\na}{n_a}			
\safemath{\nb}{n_b}			
\safemath{\coeffa}{p_i}	
\safemath{\coeffb}{q_j}	
\safemath{\seta}{\setP}		
\safemath{\setb}{\setQ}     
\safemath{\setw}{\setW}	
\safemath{\setz}{\setZ}	
\safemath{\cola}{\veca}		
\safemath{\colb}{\vecb}		
\safemath{\cold}{\vecd}		
\safemath{\inputvec}{\vecx} 	
\safemath{\error}{\vece}	
\safemath{\noiseout}{\vecz} 	
\safemath{\inputvecel}{x}
\safemath{\inputveca}{\vecx_a}
\safemath{\inputvecb}{\vecx_b}
\safemath{\outputvec}{\vecy}	
\safemath{\lambdamin}{\lambda_{\mathrm{min}}}

\safemath{\elltwo}{\ell_2}
\safemath{\ellone}{\ell_1}
\safemath{\ellzero}{\ell_0}
\safemath{\ellinf}{\ell_\infty}
\safemath{\ellinftilde}{\ell_{\widetilde\infty}}
\safemath{\licard}{Z(\coh,\coha,\cohb)}
\safemath{\xsol}{\hat{x}}
\safemath{\xbord}{x_b}		
\safemath{\xstat}{x_s}		
\safemath{\xstatLone}{\tilde{x}_s}
\safemath{\order}{\mathcal{O}} 
\safemath{\scales}{\Theta} 
\safemath{\ones}{\mathbf{1}} 
\safemath{\zeroes}{\mathbf{0}} 
\safemath{\thlone}{\kappa(\coh,\cohb)} 
\safemath{\constoneA}{\delta} 
\safemath{\constoneB}{\epsilon} 
\safemath{\nlarge}{L}				   
\safemath{\sumlarge}{S_\nlarge}
\safemath{\maxlarger}{P_\nlarge}	   
\safemath{\Pzero}{\textrm{P0}}	
\safemath{\Pone}{\textrm{P1}}
\safemath{\vecfir}{\vecw}			 
\safemath{\vecsec}{\vecz}
\safemath{\elvecfir}{w}              
\safemath{\elvecsec}{z}				 
\safemath{\nlargefir}{n}
\safemath{\normout}{\gamma}
\safemath{\auxfun}{h}
\safemath{\supp}{\textrm{supp}}

\safemath{\indexa}{\ell}
\safemath{\indexb}{r}
\safemath{\indexc}{i}
\safemath{\indexd}{j}

\safemath{\project}{P}

%% file: 0-abstract.tex
\begin{abstract}

The use of low-resolution data converters in the radio-frequency (RF) chains of all-digital massive multiple-input multiple-output (MIMO) basestations promises significant reductions in power consumption, hardware costs, and interconnect bandwidth.
\revision{We propose a quantization-aware data-detection algorithm which mitigates the performance loss of 1-bit quantized massive MIMO orthogonal frequency-division multiplexing (OFDM) systems.}
Since the system performance heavily depends on the quality of channel estimates, we also develop a nonlinear 1-bit channel estimation algorithm that builds upon the proposed data detection algorithm.  
We show that the proposed algorithms significantly outperform linear data detectors and channel estimators in terms of bit error rate. \revision{For the proposed nonlinear data detection algorithm, we develop a very large scale integration (VLSI) architecture and present implementation results on a Xilinx \mbox{Virtex-7} field programmable gate array (FPGA).}
Our implementation results are, to the best of our knowledge, the first for 1-bit massive MU-MIMO-OFDM systems and demonstrate comparable hardware efficiency  \revision{with respect to state-of-the-art linear data detectors designed for systems with high-resolution data converters, while achieving lower bit error rate.}

\end{abstract}



%% file: 1-introduction.tex
\section{Introduction}

\IEEEPARstart{M}{assive} multi-user multiple-input multiple-output (MU-MIMO) is one of the core technologies of fifth-generation (5G) wireless systems as it promises significant improvements in spectral efficiency and link reliability, compared to traditional, small-scale MIMO~ \cite{rusek14a, larsson14a}. 
In massive MU-MIMO, the infrastructure basestations (BSs) are equipped with hundreds of antenna elements that simultaneously serve tens of user equipments (UEs) in the same frequency band. 
In all-digital massive MU-MIMO basestation architectures\mbox{\cite{panagiotis20, dutta2019case}}, each radio-frequency (RF) chain is equipped with a pair of high-resolution (e.g., 10\,bit to 12\,bit) analog-to-digital converters (ADCs). 
The presence of hundreds of such high-quality RF chains inevitably results in high power consumption, interconnect data rates, and hardware costs, especially when deployed for the large bandwidths offered at millimeter-wave (mmWave) frequencies~\cite{rappaport13a, swindlehurst14a}.

To mitigate these issues, one can deploy low-resolution ADCs \cite{jacobsson17b, saxena16b, choi15a, wang14}, which is motivated by the observation that the power consumption of ADCs scales exponentially with the number of quantization bits \cite{walden99a}. 
\revision{Another benefit of deploying low-resolution ADCs is that the data rates on the fronthaul link, which connects the baseband unit and remote radio head, can be lowered significantly. 
In addition, the quality requirements on the RF circuitry (e.g. low-noise amplifiers, mixers, filters) can be relaxed, which enables further power and cost savings.}
\revision{All these benefits are attained at their greatest extent for the case of 1-bit ADCs, which can be implemented simply with 1-bit comparators and they eliminate the need for automatic gain control (AGC) circuits. This results in even more savings in RF chain power consumption and cost.}

\revision{However, due to the strong nonlinearity introduced by 1-bit ADCs, baseband processing tasks including data detection become more challenging in these systems. In this paper, we focus on data detection and channel estimation in massive MU-MIMO systems with 1-bit ADCs, operating over frequency-selective channels. However, our results can be extended to the case of multi-bit ADCs using the general framework in \cite{studer16a}, without significantly affecting the algorithm complexity. A detailed study of multi-bit case is left for future work.}

\subsection{Related Previous Work} \label{sec:PreviousWork}
Linear channel estimation and data detection algorithms for 1-bit massive MU-MIMO systems, such as maximum-ratio combining (MRC) and linear minimum mean square error (L-MMSE), have been studied in~\cite{risi14a, li16a, jacobsson17b, jacobsson15a} for systems operating in frequency-flat channels. These papers demonstrate that reliable multiuser communications is possible,  even for higher-order constellations~\cite{jacobsson15a}. Furthermore, the results in  \cite{jacobsson17b} show that 3-bit to 4-bit ADC resolution is sufficient to approach the achievable rates of infinite-resolution data converters. 

The practically more relevant case of frequency-selective channels has been studied in \cite{studer16a, mollen16c}. In \cite{mollen16c}, it has been demonstrated that linear data detection achieves acceptable performance  for wideband systems with 1-bit ADCs, assuming that the channel has a sufficiently large number of taps. For massive MU-MIMO systems with orthogonal frequency division multiplexing (OFDM), it was shown in~\cite{studer16a} that 4-bit to 6-bit ADCs are sufficient to achieve similar performance as systems infinite-resolution data converters.
While these results show that linear channel estimators and data detectors can be used in conjunction with 4-bit to 6-bit ADCs, sophisticated nonlinear channel estimation and data detection algorithms are necessary for systems that use ADCs with 3-bit or less.

To improve the performance of  1-bit  massive MU-MIMO systems, an L-MMSE channel estimator based on Bussgang's decomposition \cite{bussgang52a} has been developed in \cite{li17b}.  
Sophisticated nonlinear channel estimation and data detection algorithms have been proposed in \cite{choi15a, wang14, wen15b, wang14a, cao17, wang16, studer16a} for systems with low-resolution ADCs. The methods in \cite{wen15b, wang14a, cao17, wang16} perform data detection using generalized approximate message passing, which enables excellent error-rate performance for Rayleigh-fading channels, but at the cost of high complexity and rather poor performance in correlated or line-of-sight propagation conditions---the less complex methods in \cite{choi15a, wang14} are only suitable for frequency-flat systems.
The channel estimators and data detectors in \cite{studer16a} rely on a convex-optimization procedure, which perform well under realistic propagation conditions in coarsely-quantized massive MU-MIMO-OFDM systems but at the cost of high complexity. 
However, to the best of our knowledge, none of the above algorithms have been implemented in hardware.
 
A large number of data detector hardware designs for massive MU-MIMO systems has been proposed in the past; see, e.g.,~\cite{shahabuddin17, li17, wu16, liu20, peng17, peng20, jeon17, wu2014large}. All of these data detectors have been designed for BS architectures with high-resolution ADCs. Furthermore, these implementations are suitable for frequency-flat channels, or rely on OFDM or single-carrier frequency-division multiple access (SC-FDMA) to decompose frequency-selective channels into orthogonal, frequency-flat subcarriers. 
However, due to the severe distortion caused by 1-bit ADCs, OFDM and SC-FDMA processing does no longer result in orthogonal and frequency-flat subcarriers~\cite{studer16a}.
Hence, the use of conventional data detectors that have been designed with frequency-flat channels and high-resolution ADCs in mind inevitably result  in poor  performance in BS architectures that use 1-bit ADCs.
 
\subsection{Contributions}
We propose a new data detection algorithm and develop a corresponding VLSI design specialized for 1-bit massive MU-MIMO-OFDM systems operating over frequency-selective channels. 
Our contributions are summarized as follows:
\begin{itemize}
\item \revision{We propose a nonlinear quantization-aware data detection algorithm that solves a relaxed version of the ML detection problem in 1-bit  massive MU-MIMO-OFDM systems. The proposed algorithm includes optimizations that enable an efficient VLSI implementation.}
\item Based on our data detection algorithm, we develop a nonlinear channel estimation algorithm that mitigates the performance loss under 1-bit quantization. We further improve the quality of channel estimates using time-domain maximum likelihood estimator, which exploits correlation across subcarriers to denoise the channel estimates.
\item We use simulations to demonstrate that the proposed channel estimation and data detection methods outperform linear algorithms for frequency-selective channels.
\item \revision{We present an efficient VLSI architecture for the proposed data detection algorithm and show the first implementation results of a 1-bit massive MU-MIMO-OFDM data detector on a field programmable gate array (FPGA).}
\end{itemize}
Our simulations and FPGA implementation results show that our design achieves comparable hardware efficiency but (often significantly) lower error rate compared to data detectors that have been designed for systems with high-resolution ADCs. 

\subsection{Notation} \label{sec:notation}

Boldface lowercase and uppercase letters represent column vectors and matrices, respectively. For a matrix $\bA$, the transpose and Hermitian transpose are denoted by $\bA^\Tran$ and $\bA^\Herm$, respectively, the $k$th column is $\bma_k = [\bA]_k$, and the entry on the $m$th row and $n$th column  is $A_{m,n} = [\bA]_{m,n}$.
The $\ell_2$-norm of a vector $\bma$ and the Frobenius norm of a matrix $\bA$ are~$\|\veca\|_2$ and~$\|\bA\|_F$, respectively.
The diagonal matrix with main diagonal given by the vector $\bma$ is $\bA = \diag(\bma)$.
The $M\times N$ all-zeros and $N\times N$ identity matrices are $\bZero_{M\times N}$ and~$\bI_N$, respectively. The $N\times N$ discrete Fourier transform (DFT) matrix is denoted by $\bF$ and normalized so that $\bF\bF^H=\bI_N$.
For a vector $\bma$, the $k$th entry is denoted by $a_k = [\bma]_k$, and the real and imaginary parts   are $\Re(\bma)=\bma^R$ and $\Im(\bma)=\bma^I$, respectively. 
We use  $\boxdot$ to define an extended dot product that takes two $M\times N$ matrices $\bA$ and $\bB$ and returns an $M\times 1$ vector $\bmc = \bA\boxdot \bB$, whose $m$th element is the dot product of the $m$th rows of $\bA$ and $\bB$, i.e., $c_m = [\bA^\Tran]_m^\Tran [\bB^\Tran]_m$.
We use  $\odot$ to define a Hadamard product that takes two $M \times N$ matrices $\bA$ and $\bB$ and returns an $M \times N$ matrix $\bC = \bA \odot \bB$ whose entry on the $m$th row and $n$th column is $C_{m,n} = A^R_{m,n}B^R_{m,n} + j A^I_{m,n}B^I_{m,n}$.
\revision{The signum function $\sign(\cdot)$ operates entry-wise on vectors and for each entry~$x$ returns $+1$ if $x>0$ and $-1$ otherwise.}
A proper complex-valued zero-mean Gaussian vector $\bma$ with covariance matrix~$\mymathb{\Sigma}$ is denoted by \mbox{$\bma \sim \setC\setN(\mymathb{0}, \mymathb{\Sigma})$}.

\subsection{Paper Outline}
\revision{The rest of the paper is organized as follows. \fref{sec:SysModel} introduces the system model. \fref{sec:det_chest}  presents our  quantization-aware data detection and channel estimation algorithms for 1-bit massive MU-MIMO-OFDM systems.}
\fref{sec:VLSI} describes the VLSI architecture and shows FPGA implementation results of the data detector. \fref{sec:conclusion} concludes the paper. 

%% file: 2-system_model.tex
\section{System Model} \label{sec:SysModel}

\begin{figure}[tp]
\centering
\includegraphics[width=0.99\columnwidth]{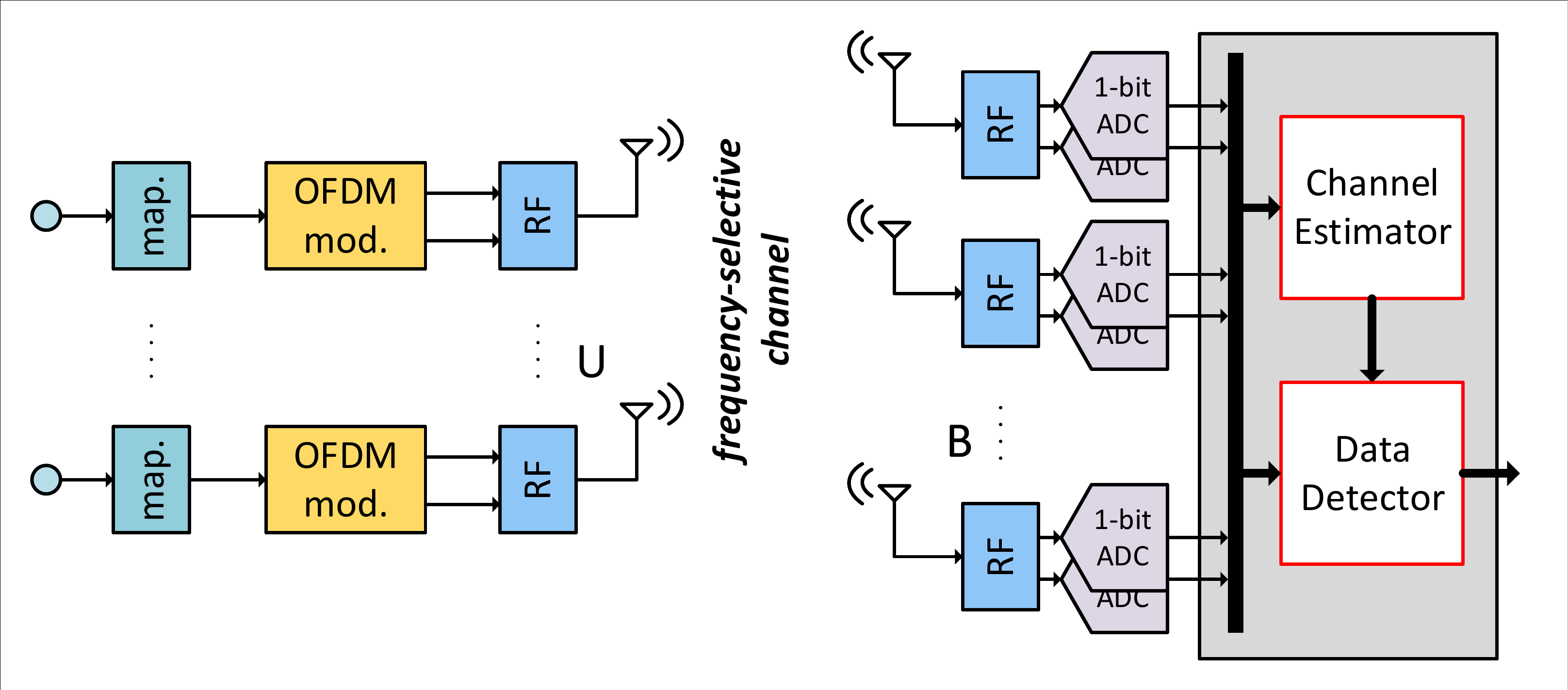}
\caption{\small Overview of a 1-bit massive MU-MIMO-OFDM uplink system. Left: $U$ user equipments (UEs) with OFDM transmission over a frequency-selective channel. Right: basestation (BS) with $B$ antennas; each BS antenna quantizes the  time-domain baseband signals with a pair of 1-bit ADCs prior to channel estimation and data detection.}
\label{fig:system_overview}
\end{figure}

We consider the uplink of a 1-bit massive MU-MIMO-OFDM system illustrated in \fref{fig:system_overview}, where $U$ single-antenna UEs communicate with a BS that is equipped with $B \gg U$ antennas.
We assume a block-fading scenario and communication over a frequency-selective channel using OFDM. 
Each OFDM symbol consists of $W = W_{\text{used}} + W_{\text{guard}}$ subcarriers, where $W_{\text{used}}$ refers to  the number of subcarriers used to carry data or pilot symbols, and $W_{\text{guard}}$ refers to the number of guard subcarriers. The set of subcarriers used for data or pilot symbols is denoted by $\Omega_{\text{used}}$ and the set of subcarriers used as guard tones is denoted by $\Omega_{\text{guard}}$.

\begin{remark}

\revision{
In what follows, we assume perfect timing and frequency synchronization at the BS. Due to the severe distortion of 1-bit quantized received signals, timing and frequency synchronization is challenging. A recent study in~\cite{jacobsson19a} has shown that accurate synchronization is feasible using the conventional Schmidl-Cox algorithm for the downlink of OFDM-based systems with 1-bit DACs. To the best of our knowledge, not much is known about uplink synchronization with 1-bit ADCs, but we expect that results in \cite{jacobsson19a} can be extended to our scenario. A detailed study of synchronization with 1-bit quantized signals is left for future work.
}

\end{remark}

The channel is modeled in the time domain by the matrices $\bH_t\in \mathbb{C}^{B\times U}$, $t= 1,\ldots,L$, where $L$ is the number of taps of the channel's impulse response.
The frequency-domain channel matrices $\bH_w\in \mathbb{C}^{B\times U}$, $w= 1,\ldots,W$, are obtained from the time-domain representation via a DFT:  
\begin{align} \label{eq:freq_domain_channel_matrices}
\bH_w\!=\! \frac{1}{\sqrt{W}}\sum_{t=1}^{L} \bH_t \exp\!\left(\!-\frac{j2\pi}{W} (t-1)(w-1-W/2) \!\right)\!,
\end{align}
noting that $\bH_t = \bZero_{B\times U}$, for $t= L+1,\ldots,W$.
To simplify notation, we often use the frequency-domain channel matrices $\bH_b\in\complexset^{W\times U}$, $b = 1,\ldots,B$, corresponding to each BS antenna. The $w$th row of $\bH_b$ is the channel vector between the $b$th BS antenna and all users on the $w$th subcarrier for all $w = 1,\ldots,W$. 
Throughout the paper, channel matrices with the subscript $b$ correspond to the $W\times U$ frequency-domain channel matrices associated with each BS antenna; channel matrices with the subscript $w$ correspond to the $B\times U$ frequency-domain channel matrices associated with each subcarrier.

\revision{Uplink communication within each channel coherence interval is divided into two phases. In the first phase, which consists of  $N_t = UT$ OFDM symbols, the UEs send pilot signals. The parameter $T$ determines the number of training symbols per UE. In the second phase, the UEs transmit $N_d$ data-carrying OFDM symbols. In what follows, we describe the  transmission model for the duration of one OFDM symbol, which is the same for both channel training and data transmission---the only difference is the choice of frequency-domain symbols.}

During the transmission of each OFDM symbol, each UE generates its own frequency-domain symbol vector $\bms_u\in \mathbb{C}^W$.
For the subcarriers reserved as guard tones, these symbols are zero, i.e., $[\bms_u]_w = 0$ for $w \in \Omega_{\text{guard}}$. 
For the other subcarriers $w \in \Omega_{\text{used}}$, these symbols are chosen from a constellation set $\mathcal{X}$ (or pilot constellation set $\mathcal{X}_t$), i.e., $[\bms_u]_w \in \mathcal{X}$ (or $[\bms_u]_w \in \mathcal{X}_t$) and are normalized as $\mathbb{E}[|[\mymathb{s}_u]_w|^2]=E_s$ for all $w$. 
Each UE then converts its frequency-domain vector $\bms_u$ into the time domain using a $W$-point inverse DFT, and transmits the resulting vector after prepending a cyclic prefix (CP) of length $P$. We assume perfect synchronization and a CP length of $P \geq L-1$, which is sufficient to avoid inter-symbol interference. 

At the BS-side, each antenna receives a noisy superposition of the UEs' signals. To simplify notation, we will use the $W\times U$ matrix $\bS$ whose $u$th column contains the frequency-domain symbols of the $u$th UE. 
Let $\bmy_b\in \mathbb{C}^W$ denote the (unquantized) signal vector received at the $b$th BS antenna after removal of the cyclic prefix. This vector can be modeled as
\begin{align} \label{eq:yb}
\bmy_b = \bF^\Herm \bmz_b + \bmn_b,
\end{align}
where $\bmn_b \sim \mathcal{CN}(\mymathb{0}_W,N_0\bI_W)$ is the thermal receive noise at the $b$th BS antenna with variance $N_0$ per complex entry, and $\bmz_b\in \mathbb{C}^W$ is the vector of noiseless frequency-domain signals associated with the $b$th BS antenna given by 
\begin{align} \label{eq:z_b}
\bmz_b = \bH_b\boxdot \bS, \quad b = 1,\ldots,B,
\end{align}
where  $\boxdot$ is the extended dot product defined in \fref{sec:notation}.

In what follows, we assume that the in-phase and quadrature baseband signals at the output of each BS RF chain  are quantized by a pair of zero-threshold 1-bit ADCs. 
For a complex-valued scalar $z$, we model this quantization operation as $r = Q(z) = \sign(z^R) + j\sign(z^I)$, which is applied element-wise to vectors and matrices. 
The $W$-dimensional vector of the 1-bit quantized observations at the $b$th BS antenna for the duration of one OFDM symbol is thus  given by
\begin{align} \label{eq:rquant}
\bmr_b = Q(\bmy_b) = Q(\bF^\Herm \bmz_b + \bmn_b).
\end{align}

Assuming $\Ex{}{\|\bH_b\|_F} = WUE_h$ for $b = 1, \ldots,B$, the average receive signal-to-noise ratio (SNR) at each BS antenna prior to quantization is given by $\rho = W_{\text{used}}UE_sE_h/(WN_0)$.

%% file: 3-data_detection_channel_estimation.tex
\section{1-Bit Data Detection and Channel Estimation} \label{sec:det_chest}

We now present our data detection and channel estimation algorithms for 1-bit massive MU-MIMO-OFDM systems, and we demonstrate their effectiveness via simulation results. 

\subsection{Quantization-Aware Data Detection with Box Constraints}\label{sec:NMLDetection}

In order to derive the quantization-aware data detection algorithm, we first formulate the ML problem and then relax its constraints to arrive at a problem that can be solved efficiently. 
We start with the likelihood of the output of a 1-bit quantizer $r=Q(\mu+n) = \sign(\mu+n)$, given the noiseless input $\mu$ and assuming that $n$ is circularly-symmetric complex Gaussian noise with variance $\No=\sigma^2$. For this model, the likelihood function is given by the following expression \cite{zymnis10a,choi15,studer16a}:
\begin{align}
p(r|\mu) & = p(r^R|\mu^R)p(r^I|\mu^I)  \notag \\
& = \Phi \!\left( \!\frac{\sqrt{2}}{\sigma}r^R\mu^R \!\right)\! \Phi\! \left( \!\frac{\sqrt{2}}{\sigma}r^I\mu^I \! \right)\!.   \label{eq:likehood_function}
\end{align}
Here, $\Phi(t) = \int_{-\infty}^{t} \frac{1}{\sqrt{2\pi}} \exp(-u^2/2) \text{d}u$, is the cumulative distribution function of a standard normal random variable. 
Now let us assume that $\bmr = Q(\bmmu + \bmn)$, where $\bmr$, $\bmmu$, and $\bmn$ are $N$-dimensional vectors, and the noise vector is distributed according to $\bmn \sim \setC\setN(\mymathb{0}, \No\bI)$.
The likelihood function of the vector $\bmr$ is given by $p(\bmr|\bmmu) = \prod_{n=1}^{N}p(r_n|\mu_n).$
Hence, the ML data detection problem corresponds to~\cite{studer16a}
\begin{align} \label{eq:ML}
\hat{\bS} = & \, \argmax_{\tilde{\bS} \in \mathbb{C}^{W\times U}} \,\prod_{b=1}^{B}p\!\left( \bmr_b|\bF^\Herm(\bH_b\boxdot \tilde{\bS})\right)\\
&\, \mathrm{subject\,\,to} \quad   [\tilde{\bS}^\Tran]_w \in \mathcal{X}^U,\qquad w \in \Omega_{\text{used}} \notag \\
& \quad \quad \quad \quad \quad\,\,\, [\tilde{\bS}^\Tran]_w = \bZero_{U\times 1},\quad  \,w \in \Omega_{\text{guard}}, \notag
\end{align}
where $\bmr_b$ is the vector of 1-bit observations at the $b$th BS antenna as defined in \fref{eq:rquant}.
\revision{Note that this problem is NP-hard, as the ML data detection problem for the infinite-resolution case, and an exhaustive search would require one to evaluate the objective function $|\mathcal{X}|^{WU}$ times.}
In order to overcome this prohibitive complexity, we relax the discrete constellation constraints. 
The same idea has been used in the special case of frequency-flat channels in \cite{wang14, choi15a, shahabuddin17, mirfarshbafan18}, and in a more general framework with multi-bit ADCs in \cite{studer16a}. 

In order to arrive at an efficient method, we relax the discrete constellation to its bounding box (convex hull), which is, for quadrature-amplitude modulation (QAM), given by 
\begin{align} \label{eq:BOX}
\mathcal{B} = \big\{x \in \mathbb{C} \mid   \max\{|x^R|, |x^I|\} \leq S_\mathcal{X} \big\},
\end{align}
where $S_\mathcal{X} = \max_{s \in \mathcal{X}} \{|s^R|, |s^I| \}$.
Concretely, we replace the constraints $[\tilde{\bS}^\Tran]_w \in \mathcal{X}^U, w \in \Omega_{\text{used}}$ in \fref{eq:ML} with $[\tilde{\bS}^\Tran]_w \in \mathcal{B}^U, w \in \Omega_{\text{used}}$. 
The resulting optimization problem is convex and can be formulated in equivalent form as follows:  
\begin{align} 
 \hat{\bS}_\text{1BOX} & =   \argmin_{\tilde{\bS} \in \mathbb{C}^{W\times U}} f(\tilde{\bS}) + \!\! \sum_{w \in \Omega_{\text{used}}} \!\!\! \mathcal{I}([\tilde{\bS}^\Tran]_w \in \mathcal{B}^U) \notag \\
& \quad \quad  + \!\!\sum_{w \in \Omega_{\text{guard}}} \!\!\!\mathcal{I}([\tilde{\bS}^\Tran]_w = \bZero_{U\times 1}). \label{eq:MLBOX}
\end{align}
Here, $f(\tilde{\bS}) = -\sum_{b=1}^{B} \log p(\bmr_b|\bF^\Herm(\bH_b\boxdot \tilde{\bS}))$ is the negative logarithm of the likelihood function in \fref{eq:ML}, and $\mathcal{I}(\cdot)$ is the indicator function which outputs zero if its input is satisfied and infinity otherwise. 
Since magnitude is lost completely in 1-bit measurements, we re-scale the output $\hat{\bS}_\text{1BOX}$ according to
\begin{align} \label{eq:normalize_detection}
\hat{\bS}^\text{norm} = \frac{E_s \sqrt{UW_{\text{used}}}}{\|\hat{\bS}_\mymath{1BOX}\|_F} \hat{\bS}_\mymath{1BOX},
\end{align}
followed by mapping the normalized output $\hat{\bS}^\text{norm}$ to the nearest constellation point in the discrete constellation set $\setX$.

\revision{The optimization problem in \fref{eq:MLBOX} can be solved using the forward-backward splitting (FBS) framework \cite{ beck2009fast,goldstein16a}. We use FBS to design an iterative algorithm that we call 1-Bit OFDM boX (1BOX) detector, which is summarized in Algorithm~\ref{alg1}.}
In each iteration of 1BOX, lines~\ref{Alg1:innloop} to~\ref{Alg1:12} calculate the (scaled) negative gradient of the objective function $f(\tilde{\bS})$ of \fref{eq:MLBOX}, and line \ref{Alg1:updateproject} increments the estimates of the current iteration in the direction of the negative gradient by a step-size of $\kappa$ and projects the result onto the box constraint set.
\revision{Calculating the gradient of $f(\tilde{\bS})$ requires the inverse Mills ratio for a normal random variable defined as}
\begin{align}\label{eq:omega}
\omega(x) = \frac{\exp(-x^2/2)}{\sqrt{2\pi}\Phi(x)},
\end{align}
\revision{and its complex version $\omega_c(z) = \omega(z^R) + j \omega(z^I)$, which is  applied entry-wise to vectors.}
With this definition, the negative gradient of the objective function $f(\tilde{\bS})$, with respect to the estimated matrix $\tilde{\bS}$, is a $W\times U$ matrix whose $w$th row is 
\begin{align}
[(\nabla f)^\Tran]_w = \frac{\sqrt{2}}{2\sigma} \bH_w^\Herm [\bV]_w.
\end{align}
Here, $\bV$ is a $B\times W$ matrix whose $b$th row is given by
\begin{align}
[\bV^\Tran]_b = \bF\! \left(\!\bmr_b \odot \omega_c\! \left(\frac{\sqrt{2}}{\sigma} \bmr_b \odot \bF^\Herm \bmz_b \right) \right)\!,
\end{align}
and  $\bmz_b$ is defined in \fref{eq:z_b}.
The quantity $\bG$ on line \ref{Alg1:11} is a scaled version of the gradient, where we omit the prefactor $\sqrt{2}/(2\sigma)$, which is absorbed into the step-size $\kappa$. 
The function $\proj_c(\cdot)$ on line \ref{Alg1:updateproject} of Algorithm \ref{alg1}, implements a projection onto the box constraints in \fref{eq:BOX}. This function operates element-wise on the entries of its input and simply performs the operation
\begin{align} \label{eq:proj}
\proj_c(x, \mathcal{B}) = \proj(x^R, \mathcal{B}) + j \proj(x^I, \mathcal{B}),
\end{align}
where $\proj(.)$ is given by $\proj(x,\mathcal{B}) = \sign(x)\min(|x|, S_\mathcal{X})$.

\begin{remark}
We note that for constant-modulus modulation schemes, such as 8-PSK, one could achieve better performance by using a circle-like polytope as the constraint set rather than a rectangular box as in \fref{eq:BOX}. However, to develop simpler VLSI implementations, we have used box constraints regardless of the modulation scheme. Projection onto more complex circle-like polytopes significantly   increases  hardware complexity~\cite{castaneda18a}.
\end{remark}

\begin{algorithm}[t]
	\begin{algorithmic}[1]
		\State \textbf{Inputs:} $\sigma$, $\{\mymathb{H}_w\}^{W}_{w=1}$, $\{\mymathb{r}_b\}^{B}_{b=1}$, $\kappa$, and $K$

		\State \textbf{Initialize:} $\mymathb{S}^{(0)} = \mymathb{0}_{U\times W}$  
		\For {$k = 1,\ldots, K$}																										\label{Alg1:outloop}
		\For {$b = 1,\ldots, B$}																											\label{Alg1:innloop}
		\State {$\bmz_b^{(k)} = \bH_b\boxdot \bS^{(k-1)}$}																				\label{Alg1:6}
		\State {$\mymathb{\alpha}^{(k)}_b = \frac{\sqrt{2}}{\sigma}\bmr_b\odot (\bF^\Herm \bmz^{(k)}_b)$}										\label{Alg1:7}
		\State {$[\bV^\Tran]_b=\bF(\bmr_b\odot \omega_c(\mymathb{\alpha}^{(k)}_b))$}\label{Alg1:8}  	
		\EndFor
		\For {$w = 1,\ldots, W$}																																\label{Alg1:10}
		\State {$[\bG^\Tran]_w = \bH^\Herm_w[\bV]_w$} 																						\label{Alg1:11}
		\EndFor																																				\label{Alg1:12}
		\State $\bS^{(k)} = \proj_c(\bS^{(k-1)} + \kappa \bG, \mathcal{B})$\label{Alg1:updateproject} 
		\EndFor
		\State \textbf{return:} $\hat{\bS}_\text{1BOX} = \mymathb{S}^{(K)}$  		
	\end{algorithmic}
	\caption{The 1BOX algorithm to solve \fref{eq:MLBOX} via FBS}
	\label{alg1}   
\end{algorithm}

\subsection{Optimization for Hardware Implementation}
The 1BOX algorithm summarized in Algorithm \ref{alg1} has two drawbacks from a hardware implementation perspective. We next introduce two solutions that address these issues.

\subsubsection{Stable Gradient Calculation}

\begin{figure}[!t]
	\centering
	\includegraphics[scale=0.6]{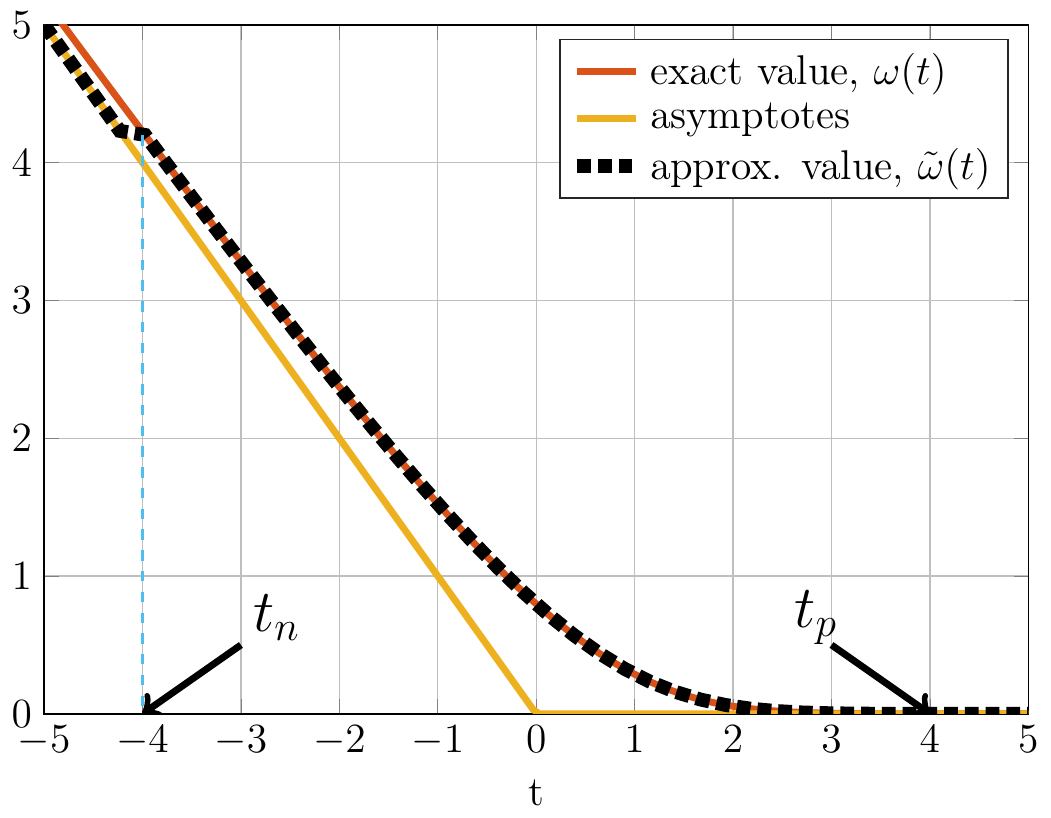}
	\caption{Mills ratio $\omega(x)$ defined in \fref{eq:omega} along with its asymptotes and approximated values used in our hardware design}
	\label{fig:GradientIllustration}
\end{figure}

\revision{The first issue arises from the inverse Mills ratio $\omega(x)$ in \fref{eq:omega}, which is numerically unstable for negative inputs of large absolute value.
Since both the nominator and denominator of \fref{eq:omega} asymptotically approach zero for large negative values of $x$, such input values result in zero-over-zero division with finite-precision arithmetic.}
In order to circumvent this issue, one can use \textit{l'H\^{o}pital}'s rule to find the negative infinity limit of $\omega(x)$, which is equal to $-x$.  For sufficiently large positive values of $x$, $\omega(x)$ produces outputs very close to zero. As illustrated in Fig. \ref{fig:GradientIllustration}, these properties of $\omega(x)$ can be exploited to simplify a hardware implementation using the approximation:
\begin{align} \label{eq: omegatilde}
\tilde{\omega}(x) = 
\left\{	\begin{array}{ll}
	0,				  			&   x\geq t_{p}\\
	\omega(x),					&  t_{n} < x < t_{p}\\
	-x,               			&   x \leq t_{n}.
	\end{array}\right.
\end{align}
The thresholds $t_{p} = 4$ and $t_{n} = -4$ have been selected based on simulations to minimize the performance degradation due to the approximation \fref{eq: omegatilde}. This approximation enables efficient hardware designs with a small look-up-table (LUT) that stores only the function values between $t_{n}$ and $t_{p}$.
	
\subsubsection{Limiting the Noise Standard Deviation} \label{sec:sigma}
The second problem arises from convergence issues at high SNR with fixed-point arithmetic. 
Adaptive step-size rules are able to deal with such convergence issues; see, e.g.,   \cite{wang14,goldstein16a}. \revision{Such rules, however, entail excessively high complexity as they require a search over suitable step sizes in every iteration that includes repeatedly evaluating the objective function (which per se requires high complexity).}
Instead, we propose a simple modification to Algorithm \ref{alg1} that simplifies our hardware design in \fref{sec:VLSI}. 
Due to the large dynamic range of the objective function at high SNR, small step-sizes are required to ensure convergence. 
From a hardware perspective, it is desirable to have a fixed step size that can be implemented with simple arithmetic shift operations. 
Since at high SNR, the role of step size is to control the dynamic range of the entries of $\bG$ to be added to $\bS^{(k-1)}$ on line \ref{Alg1:updateproject} of Algorithm~\ref{alg1}, limiting the dynamic range of the entries of $\mymathb{\alpha}_b$ on line \ref{Alg1:7} is equivalent. This effect is accomplished by thresholding the value of the noise standard deviation $\sigma$, i.e., we replace the value of $\sigma$  with $\sigma'$ whenever $\sigma < \sigma'$. Our simulations have shown that suitable values for $\sigma'$ correspond to an SNR between $10$\,dB and $15$\,dB. In \fref{sec:FTF}, we will denote the thresholded noise standard deviation by $\tilde{\sigma}$, i.e., we set $\tilde{\sigma} = \sigma'$ if $\sigma < \sigma'$ and $\tilde{\sigma} = \sigma$ otherwise.

\subsection{Linear-Quantized Data Detection} \label{sec:linear_detection}
Since linear data detection algorithms have been studied extensively in the literature for both full-resolution and low-resolution converters \cite{yin14, jacobsson17b, mollen16c}, we use them as a benchmark to evaluate the performance of the 1BOX data detector. 
Zero-forcing and L-MMSE detectors allow for efficient hardware implementations \cite{wu16, wu2014large} and achieve similar error-rate performance in massive MU-MIMO systems where $B\gg U$. 
Hence, we consider ZF detection (referred to as ZF-DET), which first converts the 1-bit received data at each BS antenna $\bmr_b$, $b=1,\ldots,B$, into the frequency domain using a DFT: $\tilde{\bmr}_b = \bF \bmr_b$.
 Let $\tilde{\bR}$ be a $W \times B$ matrix whose $b$th column is $\tilde{\bmr}_b$. Then, ZF is applied on each active subcarrier $w \in \Omega_{\text{used}}$ as follows:
\begin{align}
[\hat{\bS}^\Tran]_w = (\bH_w^\Herm \bH_w)^{-1} \bH_w^\Herm [\tilde{\bR}^\Tran]_w.
\end{align}
The outputs are then normalized as in \fref{eq:normalize_detection} and quantized to the nearest points in the constellation $\setX$.

\subsection{1-Bit Channel Estimation}
\label{sec:channelEst}

The performance of coherent data detectors depends heavily on the accuracy of the available channel estimates---this is even more critical in 1-bit quantized systems which suffer from nonlinear distortions. We next develop a method that relies on the same tools of 1BOX to calculate improved channel estimates. 
\revision{As discussed in \fref{sec:SysModel}, during the channel training phase, all UEs transmit $N_t = U T$ pilot OFDM symbols, concurrently. This results in $UTW$ training symbols in total, as each OFDM symbol contains $W$ frequency-domain symbols. The frequency-domain pilot symbols of all UEs transmitted during the $n$th pilot OFDM symbol are gathered in the matrix $\bT_n \in \mathcal{X}_t^{W \times U}$, where $\mathcal{X}_t$ is the set of pilot symbols, augmented with zero to take into account the zero-symbol used for guard subcarriers.}
\revision{To simplify exposition, we additionally introduce the per-subcarrier matrix $\bT_w \in \mathcal{X}_t^{N_t \times U}$ whose $u$th column contains the frequency-domain pilot symbols of the UE $u$ transmitted over $N_t$ pilot OFDM symbols on the $w$th subcarrier.}
Let $\bY_b$ be the $W \times N_t$ matrix associated with the $b$th BS antenna, whose $n$th column contains the unquantized time-domain samples (after removing the cyclic prefix), received during the $n$th training OFDM symbol. 
Then, the 1-bit quantized observations at the $b$th BS antenna are given by
\begin{align} \label{quantTrain}
\bR_b = Q(\bY_b) = Q(\bF^\Herm \bZ_b + \bN_b),
\end{align}
\noindent where $\bZ_b$ is a $W \times N_t$ matrix, whose $n$th column is given by $[\bZ_b]_n = \bH_b \boxdot \bT_n$, and the entries of $\bN_b$ are i.i.d.\ circularly-symmetric complex Gaussian with variance $N_0$. 
Our objective is to obtain channel estimates $\{\hat{\bH}_w\}_{w=1}^W$ based on the 1-bit observations $\{\bR_b\}_{b=1}^{B}$ and the known pilot matrices $\{\bT_n\}_{n=1}^{N_t}$.

\subsubsection{Linear-Quantized Channel Estimation} \label{sec:lin-quant-chest}
A na\"ive way for obtaining channel estimates from 1-bit measurements is to ignore quantization altogether and perform $W$ independent channel estimation tasks relying on the orthogonality of OFDM together with linear channel estimators, such as ZF or L-MMSE, on a per-subcarrier basis \cite{studer16a, mollen16c}.
Since ZF and L-MMSE channel estimation provides similar performance in  massive MU-MIMO, we focus on ZF channel estimation.
Similar to ZF-DET,  ZF channel estimation (called ZF-CHEST) first converts the 1-bit measurements at each BS antenna $\bR_b$ into the frequency domain according to $\tilde{\bR}_b = \bF \bR_b$, $b = 1,\ldots,B$. Then, the channel estimates for each BS antenna $b$ and for each active subcarrier $w \in \Omega_{\text{used}}$ are obtained as follows:
\begin{align}
[(\hat{\bH}_w)^\Tran]_b = (\bT_w^\Herm \bT_w)^{-1}\bT_w^\Herm [\tilde{\bR}_b ^\Tran]_w.
\end{align}

We note that a Bussgang-based L-MMSE (BL-MMSE) channel estimator has been proposed in~\cite{li17b} that achieves superior performance than the conventional L-MMSE channel estimator by taking into account the nonlinearities caused by 1-bit ADCs. A similar approach can be used to design BL-MMSE-based data detectors \cite{abdallah18, nguyen19}. 
Unfortunately, the complexity of these methods is prohibitive in OFDM systems, as they require the inversion of a $UW \times UW$ matrix, caused by the fact that 1-bit quantization destroys orthogonality that enables one to decouple the estimation problem into independent problems for each subcarrier. 
Furthermore, the matrix to be inverted applies a nonlinear arcsine function to each entry, which prevents efficient  iterative methods to find the inverse.

\begin{figure*}[tp]
	\centering
	\subfigure[$B=128$, $U=8$, 8-PSK]{\includegraphics[width=0.4\textwidth]{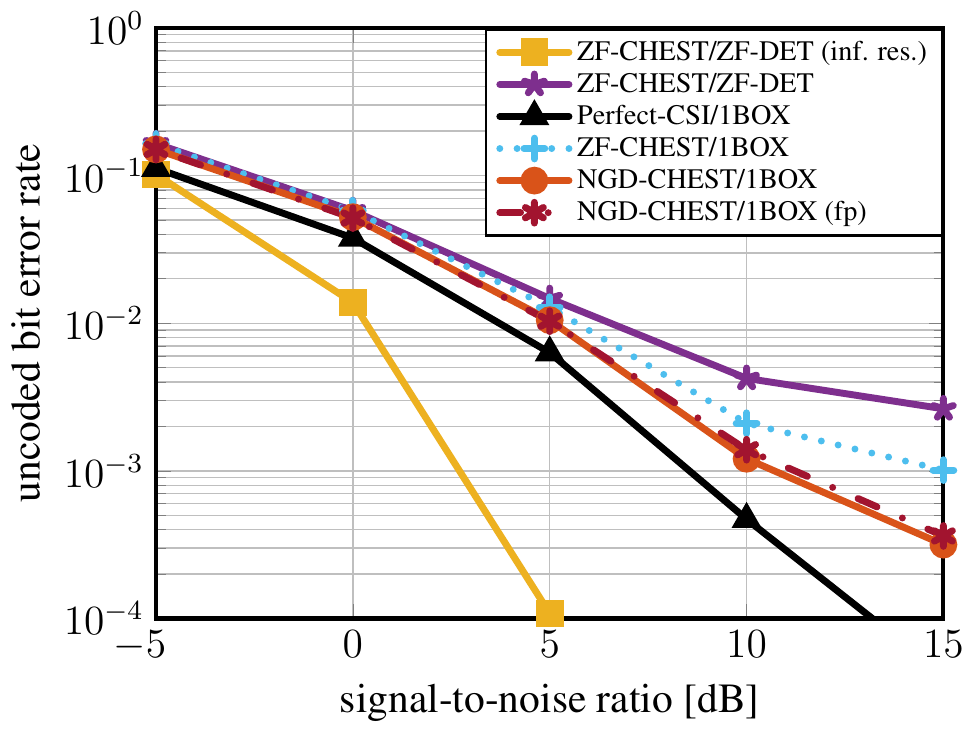}}
	\hspace{1.2cm}
	\subfigure[$B=128$, $U=8$, 16-QAM]{\includegraphics[width=0.4\textwidth]{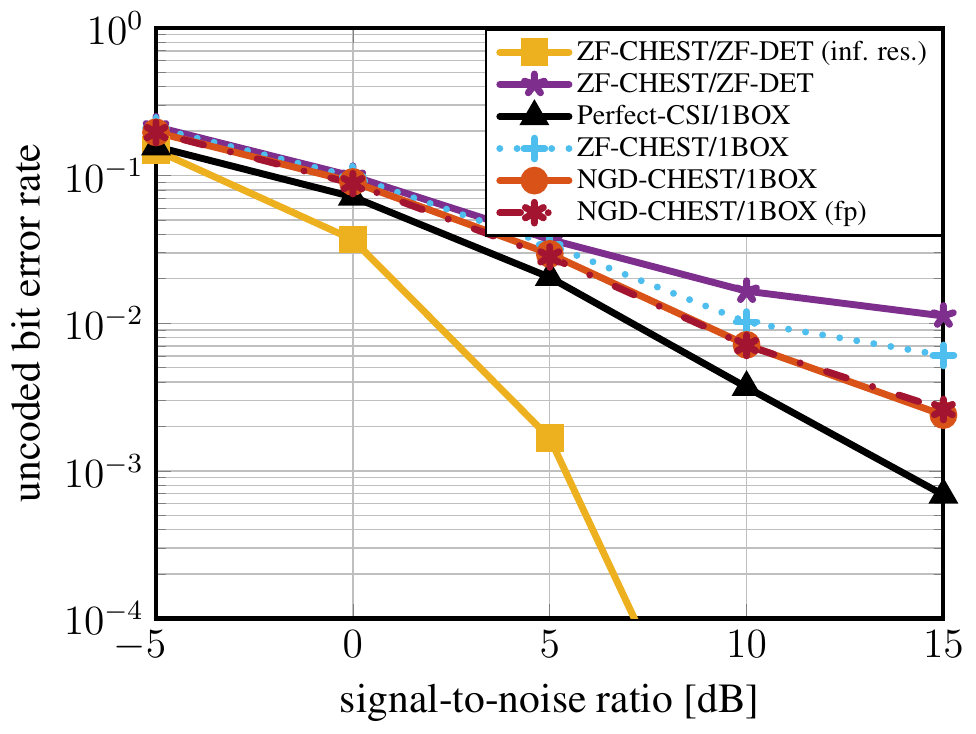}}\\	
	\subfigure[$B=64$, $U=4$, 8-PSK]{\includegraphics[width=0.4\textwidth]{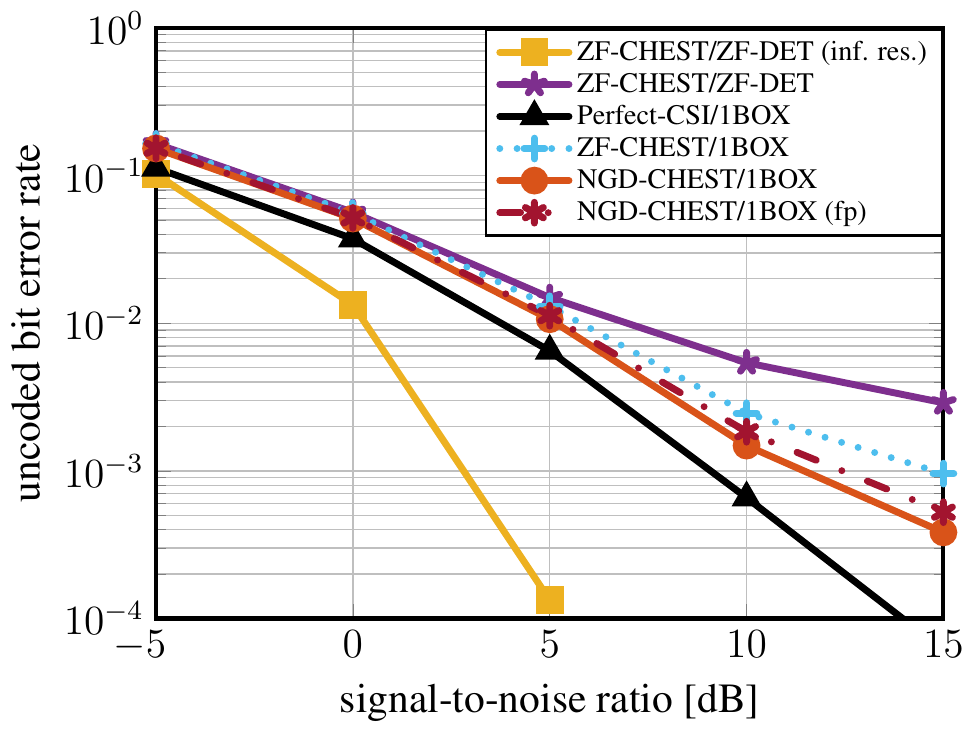}}
	\hspace{1.2cm}
	\subfigure[$B=64$, $U=4$, 16-QAM]{\includegraphics[width=0.4\textwidth]{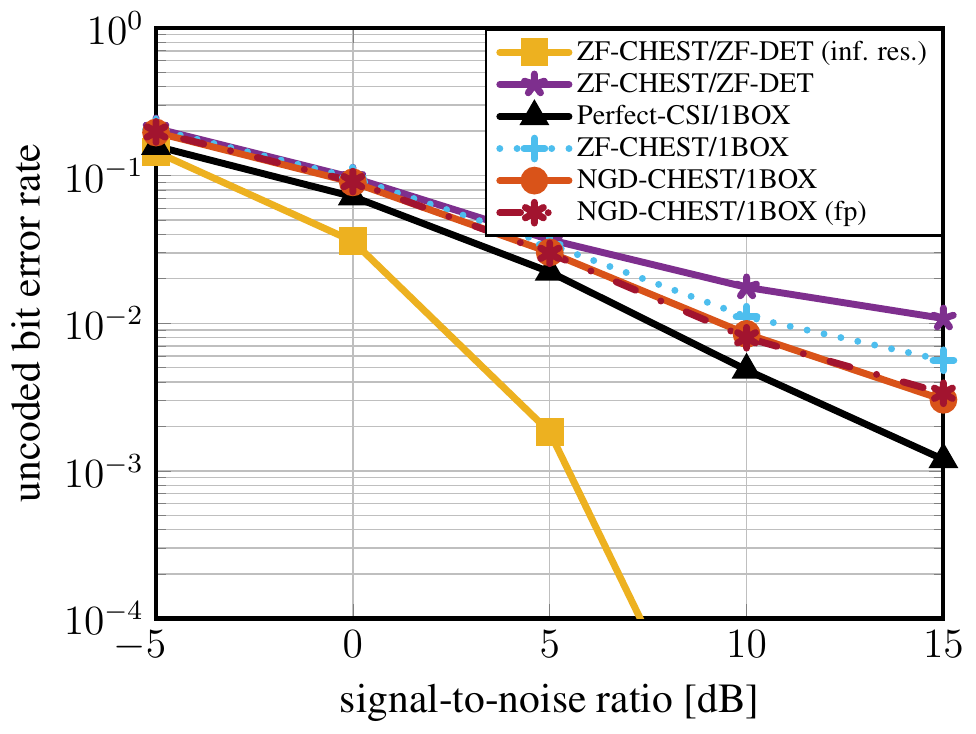}}
	\caption{Uncoded bit error rate (BER) of the proposed 1BOX detector  with perfect-CSI and channel estimates obtained from ZF-CHEST or NGD-CHEST. As a reference, the BER performance of ZF detector with ZF-CHEST, both with 1-bit and infinite resolution ADCs is also included.}
	\label{fig:BER}
\end{figure*}

\subsubsection{ML-based Channel Estimation} \label{sec:NGD-CHEST}
Another approach to obtain improved channel estimates   $\hat{\bH}_b$, $b = 1, \ldots, B$, is to directly solve the ML channel estimation problem
\begin{align} \label{eq:MLCHEST}
&\hat{\bH}_b = \argmax_{\tilde{\bH}_b \in \mathbb{C}^{W\times U}} \prod_{n=1}^{N_t}p([\bR_b]_n|\bF^\Herm(\bT_n\boxdot \tilde{\bH}_b)),
\end{align}
which is derived analogously to \fref{eq:ML}.
The idea of ML-based channel estimation with 1-bit quantized measurements for frequency-flat channels has been put forward in \cite{choi15a}. We emphasize, however, that the method in \cite{choi15a} is not directly applicable to OFDM systems, and the general ML problem needs to be derived for such systems, as done in \fref{eq:MLCHEST}.
The problem in \fref{eq:MLCHEST} is convex and we propose to solve it using normalized gradient descent (NGD), which we call NGD-CHEST. 
The algorithm resembles that of the 1BOX data detector in \fref{eq:ML}, except that (i) there are no constraints on the channel matrices (and hence no projection operation required), (ii) we assume that the pilot matrices $\bT_n$, $n = 1,\ldots, N_t$, are known, (iii) we initialize the algorithm with the ZF-CHEST estimates to improve convergence.
\revision{
Due to the structural similarity between 1BOX and NGD-CHEST, it would be possible,  with only a few modifications, to use the same VLSI architecture proposed for 1BOX in \fref{sec:VLSI} to also carry out the NGD-CHEST algorithm. However, due to the short time available for channel training in each channel coherence interval, it would not be practical to carry out both NGD-CHEST and 1BOX data detection with the same hardware instance---separate instances should be used for each task.
}
Next, we describe channel denoising and normalization techniques that are applied to NGD-CHEST and ZF-CHEST outputs to improve the channel estimates.
 
\subsubsection{Channel Denoising and Normalization} \label{sec:channel_denoising}
 
In most practical systems, the number $L$ of channel taps in the time domain is less than the number of OFDM subcarriers---this implies that adjacent subcarriers are correlated. 
One can exploit this property to denoise the channel estimates by means of a time-domain maximum likelihood estimator (TDMLE) \cite{haene08}. For the frequency-domain channel $\hat{\bmh}_{b,u} \in \mathbb{C}^{W_{\text{used}} \times 1}$ between the $b$th BS antenna and $u$th user, a denoised channel estimate can be obtained as follows:
\begin{align} \label{eq:denoise}
\hat{\bmh}^{\text{denoised}}_{b,u} = \bF_{\Omega_{\text{used}}} (\bF_{\Omega_{\text{used}}}^\Herm \bF_{\Omega_{\text{used}}})^{-1} \bF_{\Omega_{\text{used}}}^\Herm \hat{\bmh}_{b,u}.
\end{align}
Here, $\bF_{\Omega_{\text{used}}}$ is a $W_{\text{used}} \times L$ matrix, constructed from a $W$-point DFT matrix by taking its first $L$ columns and the rows indexed by $\Omega_{\text{used}}$. The resulting denoised channel estimates are collected in the matrices $\hat{\bH}^{\text{denoised}}_b$, $b = 1,\ldots, B$.

Since for 1-bit quantizers amplitude information is lost completely, we apply a re-scaling procedure analogous to \fref{eq:normalize_detection}. More concretely, we assume that the BS is able to acquire an estimate of the average channel gain at each BS antenna $\gamma_b = \Ex{}{\|\bH_b\|_F}$.\footnote{\revision{Such information can, for example, be acquired by using one or multiple higher-resolution ADCs during a training phase.}} Due to the spatial proximity of BS antennas, we assume that $\gamma_b = \gamma$ for all $b = 1,\ldots, B$. In order to minimize the mean squared error (MSE) between the exact channel matrices $\{\bH_b\}_{b=1}^B$ and their estimates $\{\hat{\bH}_b\}_{b=1}^B$, the BS re-scales the channel estimates as follows:
\begin{align} \label{eq:normalize_chest}
\hat{\bH}^{\text{norm}}_{b} = \frac{\gamma}{\| \hat{\bH}^{\text{denoised}}_{b} \|_F} \hat{\bH}^{\text{denoised}}_{b}, \quad b=1,\ldots,B.
\end{align}

\subsection{Complexity Analysis} \label{sec:complexity}
\revisionS{In this section, we provide analytic complexity expressions for the 1BOX and NGD-CHEST algorithms, measured in terms of the number of real-valued multiplications. 
Each iteration of 1BOX, as shown in Algorithm~\ref{alg1}, consists of two for-loops with $B$ and $W$ iterations. 
Line~\ref{Alg1:6} of Algorithm~\ref{alg1} involves $W$ inner products between $U$-entry vectors. By assuming that each complex-valued multiplication requires four real-valued multiplications, this line requires $4UW$ real-valued multiplications. The computations on line~\ref{Alg1:7} and \ref{Alg1:8} involve DFT operations on $W$-entry vectors that can be implemented using fast Fourier transforms (FFT). Each $W$-point FFT requires $2W \log_2W$ real-valued multiplications, assuming a radix-2 implementation. The complexity of calculating $\omega_c$  for a scalar input using look-up-tables can be approximated by one real-valued multiplication. The Hadamard products with the vectors $\bmr_b$ on lines \ref{Alg1:7} and \ref{Alg1:8} do not require actual multiplications and can be implemented efficiently with conditional negations as described in \fref{sec:FTF}. Line~\ref{Alg1:11} consists of the product of a $U \times B$ matrix by a $B \times 1$ vector, which requires $4UB$ real-valued multiplications. Multiplications with the real-valued constant $\kappa$ of all entries of $\bG$ on line~\ref{Alg1:updateproject}, requires $2UW$ real-valued multiplications. In total, 1BOX involves $8BUW + 4BW\log_2W + BW + 2UW$ real-valued multiplications per algorithm iteration.}

\revisionS{To obtain an estimate of each of the $W \times U$ channel matrices $\bH_b$, $b = 1,2, \ldots,B$, we need to run NGD-CHEST, which is similar to 1BOX, with the exception that the dimension $B$ is replaced by~$N_t$. Therefore, the overall complexity of obtaining estimates for $B$ channel matrices is given by $BK(8N_tUW + 4N_tW\log_2W + N_tW + 2UW)$ for $K$ algorithm iterations.}

\subsection{Simulation Results}
\label{sec:SimRes}

\revision{To demonstrate the effectiveness of NGD-CHEST and the 1BOX data detector, we now present simulation results and a comparison with linear channel estimators and data detectors designed to operate with high-resolution ADCs}. 

\subsubsection{Simulation Settings}
We consider a 1-bit massive MU-MIMO-OFDM system with $W = 128$ subcarriers, whose middle part of $W_{\text{used}} = 100$ subcarriers are used for data and pilots, and the remaining $W_{\text{guard}} = 28$ subcarriers at the two sides are left unused.
During the channel estimation phase, all UEs simultaneously transmit pilots.  
Since non-sparse pilot matrices perform better than diagonal training matrices, a phenomenon that has been observed  in~\cite{studer16a}, we set the frequency-domain pilot matrices $\bT_n$, $n=1,\ldots,N_t$, to contain random QPSK symbols that are known to the BS. 
In our simulations, we use $N_t = UT$ training OFDM symbols with $T = 2$.
For NGD-CHEST, we set the number of algorithm iterations to $K = 5$ with a fixed step-size of $1/16$.
For both ZF-CHEST and NGD-CHEST, we use TDMLE channel denoising and normalization techniques outlined in \fref{sec:channel_denoising}, to improve the channel estimates.
During the data transmission phase, the UEs generate frequency-domain symbols from either 8-PSK or 16-QAM constellations. 
For the 1BOX detector, we use $K = 3$ iterations and set the step-size to $\kappa = \sqrt{2}/64$ for the floating-point experiments and $\kappa = 1/32$ for our fixed-point results as we absorb the $\sqrt{2}$ factor into the scaling schedule of the FFT hardware design. In addition, the choice of $\kappa = 1/32$ simplifies fixed point design as it can be implemented with trivial arithmetic right shifts.

\subsubsection{Error-Rate Performance}
\fref{fig:BER} shows uncoded bit error rate (BER) results for (i) a systems with $B = 128$ BS antennas and $U = 8$ UEs and (ii) a systems with $B = 64$ BS antennas and $U = 4$ UEs. For both systems, we use Gray-coded 8-PSK and 16-QAM. Each plot in  \fref{fig:BER}, contains six curves: (i) ZF-CHEST followed by ZF-DET, with infinite resolution ADCs for both channel estimation and data detection (used as a reference), (ii) 1-bit ZF-CHEST followed by 1-bit ZF-DET, (iii) 1BOX detector with perfect CSI, (iv) 1-bit ZF-CHEST followed by 1BOX detection, (v) NGD-CHEST followed by 1BOX detection and (iv) NGD-CHEST with the fixed-point version of 1BOX detection (denoted by ``(fp)'' on the figure legends) which uses the fixed-point implementation parameters detailed in \fref{sec:fixedpoint}.
As we can see from \fref{fig:BER}, the 1BOX detector combined with NGD-CHEST achieves significantly better error-rate performance compared to linear quantized uplink processing, i.e., ZF-DET with ZF-CHEST. The SNR gap at $\textit{BER} = 10^{-2}$  ranges from $1.5$\,dB for the $128 \times 8$ setup with $8$-PSK signaling to $8$\,dB in the $128 \times 8$ setup with $16$-QAM signaling. 
Additionally, we see that obtaining channel estimates from NGD-CHEST results in significantly better error-rate performance than those from ZF-CHEST.

\revision{In  \fref{fig:BER}, we observe that the performance of the fixed-point version of 1BOX that corresponds to the hardware implementation detailed in \fref{sec:VLSI}, closely match those of  floating-point performance.}
\revision{Finally, we observe that the proposed NGD-CHEST and 1BOX algorithms, need higher SNRs to achieve the same BER as that of ZF-CHEST and ZF-DET for high-resolution systems. This indicates the existence of a trade-off between the error-rate performance and the reduction in cost and power consumption of basestation RF chains, when using 1-bit ADCs. A detailed study of this trade-off and an assessment of the overall system cost and power consumption as a function of ADC resolution is left for future work.}

\begin{remark}

\revision{A key limitation of data detection with 1-bit ADCs is that supporting higher-order modulation schemes (such as 64-QAM or higher) is challenging \cite{studer16a}. While the proposed algorithm shows acceptable performance with 8-PSK and 16-QAM, it does not perform well for higher-order modulations schemes, especially for SNR values typically encountered in mmWave systems. Therefore, 1-bit data detection is suitable for systems that operate at lower per-user data rates in exchange for reduced RF chain cost and power consumption.}

\end{remark}

%% file: 4-VLSI.tex

\begin{figure*}[tp]
	\centering
	\includegraphics[width=0.98\textwidth]{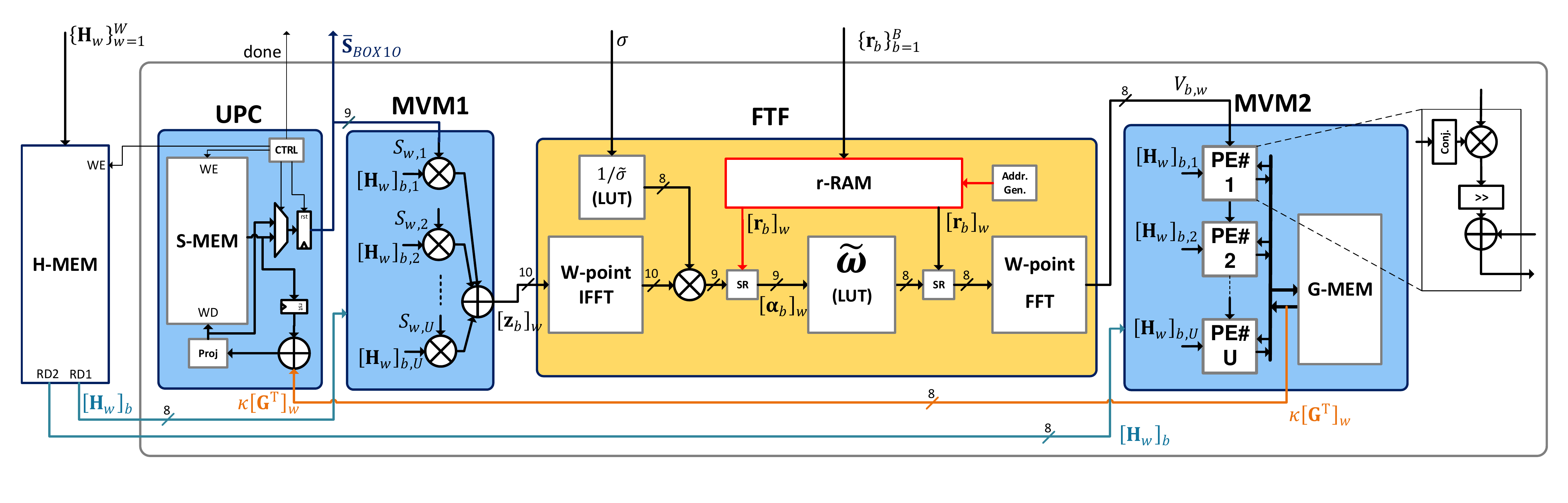}
	\vspace{-0.35cm}	
	\caption{\small VLSI architecture  of the 1BOX data detection algorithm for 1-bit massive MU-MIMO-OFDM systems.}
	\label{fig:Architecture}
\end{figure*}

\section{Architecture and FPGA Implementation} \label{sec:VLSI}
We now  present a VLSI architecture for the 1BOX data detection algorithm and show reference FPGA implementation results. We then provide a comparison with existing linear data detectors that have been developed for infinite-resolution massive MU-MIMO systems.
To the best of our knowledge, this  is the first data detector implementation for 1-bit massive MU-MIMO systems reported in the open literature.

\subsection{Architecture Overview and Operation Principles} \label{sec:VLSIArch}
The proposed VLSI architecture is shown in \fref{fig:Architecture} and consists of the following five modules: 
(i) UPC (short for update, project, and control), (ii) MVM1 (short for matrix-vector multiplication unit~1), (iii) FTF (short for frequency-time-frequency),  (iv) MVM2 (short for matrix-vector multiplication unit~2), and (v) H-MEM (short for $\bH$-memory).
The UPC module carries out the operations on line \ref{Alg1:updateproject} of Algorithm~\ref{alg1}, as well as preparing the control signals;
the MVM1 module is responsible for the matrix-vector multiplication on line~\ref{Alg1:6};
the FTF module performs the operations on lines~\ref{Alg1:7} and~\ref{Alg1:8};  
the MVM2 module is responsible for the matrix-vector multiplication on line \ref{Alg1:11}; and 
the H-MEM module stores the frequency-domain channel matrices $\{\hat{\bH}_w\}_{w=1}^W$. 
The signal names in \fref{fig:Architecture} correspond to the variables in Algorithm~\ref{alg1}, e.g., $[\bH_w]_b$ represents the channel vector of the $b$th BS antenna over the $w$th subcarrier. 

\revision{The proposed architecture has been optimized in terms of hardware efficiency, measured in throughput per FPGA resources.}
\revision{With this optimization strategy,  our goal is to minimize the resource consumption needed to achieve a specific throughput. Therefore, even though each instance of the proposed architecture may achieve relatively low throughput, it is possible to scale up the throughput by replication, i.e., by instantiating $N$ parallel designs that process different sets of receive signals. This approach enables us to increase the throughput by a factor of $N$ while maintaining the same hardware efficiency.}

\begin{figure*}[tp]
	\centering
	\includegraphics[width=0.85\textwidth]{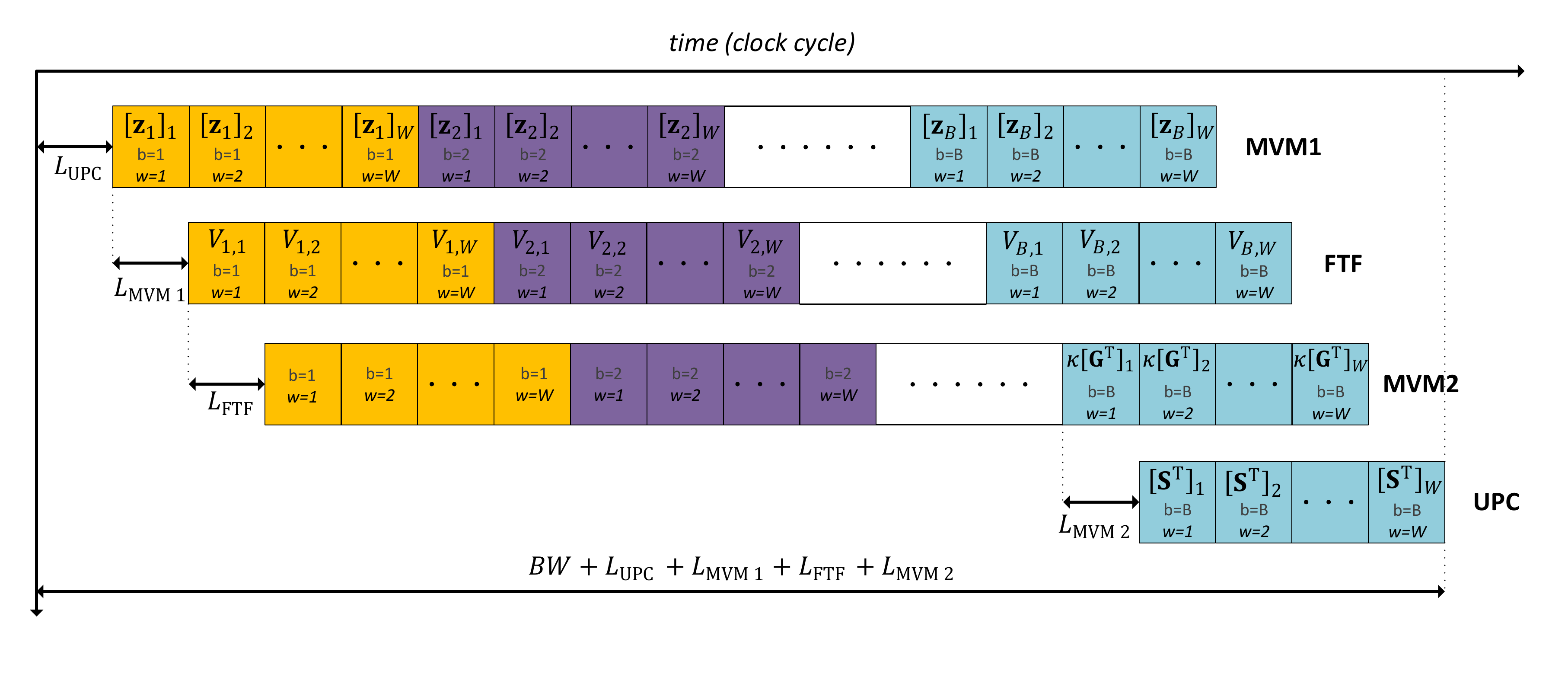}
	\vspace{-0.8cm}	
	\caption{\small Sequence of operations carried out by the submodules of the proposed architecture within one iteration of 1BOX.}
	\label{fig:timeline}
\end{figure*}

\revision{In addition, the proposed architecture operates in streaming fashion, which reduces the overhead of control and data buffering. 
As illustrated in \fref{fig:timeline}, the architecture processes the data of the $b$th BS antenna over $W$ consecutive clock cycles---one clock cycle for each of the $W$ subcarriers. For example, the elements of the $W \times 1$ vector $\bmz_b^{(k)}$ on line~\ref{Alg1:6} of Algorithm~\ref{alg1}, are produced in MVM1 sequentially over $W$ consecutive clock cycles. Consequently, it takes $BW$ clock cycles to compute all vectors $\bmz_b^{(k)}$, for $b = 1,2,\ldots,B$, in the $k$th iteration of 1BOX.
Therefore, each algorithm iteration requires $D = BW + L_{\text{UPC}} + L_{\text{MVM1}} + L_{\text{FTF}} + L_{\text{MVM2}}$ clock cycles, where $L_{\text{X}}$ denotes the latency of module ``X,'' caused by pipelining.}
\revision{Consequently, our data detector architecture achieves a  sustained throughput of}
\begin{align}
\Theta = \frac{|\mathcal{X}|UW_{\text{used}}f}{KD} \quad \text{[Mb/s]},
\end{align} 
where $|\mathcal{X}|$ is cardinality of the constellation set and $f$ is the circuit's clock frequency in MHz. 
\revision{A detailed discussion of the timing schedule is provided in \fref{sec:timing}.}

\subsection{Architecture Details } \label{sec:arch_details}
The operating principles and implementation details of the five  modules  are detailed next.

\subsubsection{UPC} 
This module contains a $U\times W$ memory block labeled S-MEM that stores the matrix of estimated symbols~$\bS^{(k)}$ at iteration $k$ . 
All entries of this memory block are updated at the end of each iteration with the values obtained from the MVM2 module. 
The control unit, labeled ``CTRL'', is responsible for synchronizing operations of the entire architecture, as all other modules mainly consist of steaming data paths. 
Since the matrix $\bS^{(k)}$ accumulates the estimates over iterations as in Algorithm~\ref{alg1}, it has to be initialized at the beginning of the first iteration. In order to avoid wasting any clock cycles for erasing the content of the S-MEM block and to allow for continuous processing, the control unit asserts the reset signal of the output register of the UPC module, so that initial values of zero are provided to the next module during the first $W$ clock cycles of the first iteration.

At the end of each iteration, the results of the MVM2 module are ready in G-MEM. The UPC module receives the content of G-MEM during the last $W$ clock cycles of each iteration. In the $w$th cycle, UPC receives the vector $\kappa[\bG^\Tran]_w$, retrieves the $w$th column of S-MEM which contains $[\bS^\Tran]_w$, adds it with $\kappa[\bG^\Tran]_w$, and writes back the results to $w$th column of S-MEM after applying the projection operation, as shown on line \ref{Alg1:updateproject} of Algorithm \ref{alg1}. 
When updating the S-MEM block with the result from the MVM2 module during the first iteration, the control unit asserts the reset signal of the register before the accumulator to ensure that the S-MEM content from the previous detection task does not get accumulated. In addition, during the first $W$ clock cycles of each iteration, the control unit sets the select signal of the multiplexer (MUX) at the output of the UPC module so that its output comes directly from the projection operation, since during these clock cycles the content of S-MEM is being updated and hence not ready to be passed to the output.

\subsubsection{MVM1} 
This module consists of $U$ complex-valued multipliers and a balanced adder tree that sums the results of $U$  multipliers. In every clock cycle, two $U$-dimensional vectors enter the module and their inner product is computed after  $L_{\text{MVM1}}$ clock cycles, including additional clock cycles caused by pipelining. The value of $L_{\text{MVM1}}$ depends on the number of UEs $U$. For $U = 4$ and $U = 8$, for example, the number of clock cycles are $4$ and $5$, respectively.

\subsubsection{FTF} \label{sec:FTF}
The FTF module contains the FFT and inverse FFT (IFFT) submodules, which are implemented using the Xilinx LogiCORE FFT IP with radix-2 \textit{pipelined streaming I/O} architecture. 
This streaming FFT architecture achieves high throughput and provides continuous processing capability, i.e., accepts one sample of a $W$-point vector per clock cycle and produces one entry of the resulting transform per clock cycle, after a latency of $L_{\text{FFT}}$ clock cycles. 
The IFFT and FFT submodules carry out the operations on lines \ref{Alg1:7} and \ref{Alg1:8} of Algorithm \ref{alg1}, respectively. 
Each output of the IFFT core, is multiplied by $1/\tilde{\sigma}$ (cf. \fref{sec:sigma}) and then sign-refined\footnote{Sign-refinement of a complex signal $x$ with a complex 1-bit signal~$r$ refers to the operation $x \odot r$; each real and imaginary part of $x$ is negated if the corresponding part of $r$ is $-1$ and stays unaltered otherwise.} 
by the corresponding 1-bit received signal $[\bmr_b]_w$ (carried out with the logic labeled with ``SR'' on \fref{fig:Architecture}) to produce the vector $\mymathb{\alpha}_b$ as shown on line \ref{Alg1:7}. Note that the missing factor of $\sqrt{2}$ is absorbed to the scaling schedule of the IFFT implementation. 
The result, which is in the time domain, is then fed to the nonlinear function $\tilde{\omega}$ defined in \fref{eq: omegatilde}, which is implemented by a small LUT, as detailed in \fref{sec:NMLDetection}. 
The output of the $\tilde{\omega}$ submodule is once again sign-refined before being converted back to the frequency-domain by the FFT core. For an OFDM system with $W = 128$ subcarriers, the FTF module has a latency of $L_{\text{FTF}} = 702$ clock cycles, which is the sum of the latencies of all submodules.

\begin{table*}[tp]
	\centering
	\renewcommand{\arraystretch}{1.1}
	\begin{minipage}[c]{1\textwidth}
		\vspace{-0.1cm}
		\centering
		\caption{Implementation results and comparison on a Xilinx Virtex-7 XC7VX690T FPGA }
		\label{tbl:implresults}
\resizebox{0.98\textwidth}{!}{%
		\begin{tabular}{@{}lccccccc@{}}
			\toprule
			
			Detector  & 1BOX  & 1BOX & NS-SCFDMA \cite{wu2014large} & OCD \cite{wu16} & PGS \cite{wu17} & RCG \cite{liu20}  & ESDBB \cite{tan20} \\
			System dimension ($B\times U$) & $64\times 4$ & $128\times 8$ & $128\times 8$ & $128\times 8$ & $128\times 8$ & $128\times 8$ & $128\times 8$\\
			Designed for 1-bit ADCs?& \bf yes & \bf  yes & no & no & no & no & no  \\
			OFDM or SC-FDMA & OFDM & OFDM & SC-FMDA  & no & no & no & no \\
			\midrule
			{Slices} & 1\,201  (1.11\%) & 1\,627  (1.5\%) & 48\,244  (45\%) & 11\,094  (10\%) & NA\footnote{\label{hh}Number not reported.} & NA\footref{hh}  & NA\footref{hh}  \\
			{LUTs} & 3\,125 (0.72\%) & 3\,952 (0.91\%) & 148\,797 (34\%) & 23\,914 (5.5\%) & 10\,085 (2.3\%) & 4\,587 (1.0\%) & 3\,631 (0.83\%)\\
			{FFs} & 4\,412 (0.51\%) & 5\,221 (0.6\%)& 161\,934 (19\%) & 43\,008 (4.96\%) & 12\,074 (1.4\%) & 18\,782 (2.1\%) & 7\,515 (0.85\%) \\
			{DSP48 units} & 52 (1.44\%) & 84 (2.33\%) & 1\,016 (28\%) & 774 (21.5\%) & 266 (7.4\%) & 972 (27.0\%) & 272 (7.42\%)\\
			{Block RAMs} & 22.5  & 72  & 16 & 2 & 4 & NA\footref{hh}  & NA\footref{hh} \\
			\midrule
			Max.\ clock frequency [MHz] & 303 & 303 & 317 & 258 & 322 & 210 & 210\\
			{Latency [clock cycles]} & 26\,706 & 51\,282 & 196 & 795 & 204 & NA & 322\\
			{Throughput [Mb/s]} & 18.14 \footnote{\label{ff}The throughput of the 1BOX detector is reported for $K = 3$ iterations and 16-QAM.}  & 18.89 \footref{ff}  & 414 \footnote{\label{sf}The throughput of the design has been scaled to 16-QAM.} & 250 \footref{sf} & 51 \footref{sf} & 420 \footref{sf} & 31.3\\
			\midrule
			{Throughput/LUT}  & {5\,808} & {4\,784} & {2\,782} & {10\,398}  & {5\,024} & {91\,563} & {8\,619} \\
			{Throughput/NRC \footnote{Normalized Resource Consumption (NRC) is calculated as $\text{NRC} = \text{LUT} + \text{FF} + 280 \times \text{DSP48}$ \cite{chen16}.}}  & {820} & {578} & {695} & {881}  & {527} & {1\,421} & {358} \\
			\bottomrule
		\end{tabular}
	}		
	\end{minipage}

\end{table*}

\subsubsection{MVM2} 
This module is responsible for the matrix-vector multiplications on line \ref{Alg1:11} of Algorithm \ref{alg1}, as well as the multiplication by the step size $\kappa$, i.e., computes $\kappa \bG$. 
To this end, the MVM2 module contains $U$ processing elements (PEs), which consists of a complex-valued multiplier, a complex-valued adder, and logic to perform complex conjugation (denoted by ``conj.'') and shifting to carry out the multiplication by the step size ($\kappa$), since the step size is chosen to be a negative power of two (i.e. 1/32). The architecture details of a PE is shown on the right side of \fref{fig:Architecture}.

We note that the sequence of operations carried out by the MVM2 module is not exactly the same as shown by the for-loop between lines \ref{Alg1:10} to \ref{Alg1:12} of Algorithm \ref{alg1}; this is because the MVM2 module receives the elements of the $B\times W$ matrix $\bV$, defined on line \ref{Alg1:8} of Algorithm \ref{alg1}, in a row-by-row fashion, over $BW$ clock cycles, while the line \ref{Alg1:11} of the algorithm shows the multiplication with one column of $\bV$ at a time.
The module receives the $b$th row $[\bV^T]_b$, during clock cycles $(b-1)W+1$ to $bW$. As soon as it receives the $b$th element of $[\bV]_w$ (which happens $W$ clock cycles after receiving the $(b-1)$th element of $[\bV]_w$), the $U$ PEs multiply it with the $b$th column of $\bH_w^\Herm$, right-shift the product by $\log(\kappa^{-1})$ bits and accumulate the result in the $w$th column of G-MEM. This process continues until all $BW$ elements of $\bV$ have been received and the $w$th column of G-MEM contains the result of the matrix-vector multiplication $\bH_w^\Herm [\bV]_w$, for $w = 1,...,W$. 	The MVM2 module has a latency of $L_{\text{MVM2}} = 4$, due to its internal pipeline stages.

\subsubsection{H-MEM} 
This module is a $BW \times U$ memory that stores the $W$ frequency-domain  channel matrices $\{\hat{\bH}_w\}^{W}_{w=1}$. During the last iteration of each data detection task, in which the entries of the channel matrices are used for the last time, the control unit within the UPC module signals the channel estimation module that the H-MEM is ready to receive new matrices pertaining to the next data detection problem and asserts the write-enable (WE) signal of the H-MEM block. 

\begin{remark}
\revision{
All modules of the proposed architecture have been carefully pipelined in order to improve hardware efficiency. As a result, our designs achieve relatively high clock frequencies. We used pipelining registers at the inputs and outputs of each multiplier, implemented by DSP48 units of the FPGA. However, in order to keep our architecture diagrams simple, we do not show these pipelining registers in \fref{fig:Architecture}. The critical path of the design is shown with red color in  \fref{fig:Architecture} and passes through the  memory array ``r-RAM'' within the FTF module.}
\end{remark}

\revision{\subsection{Timing Schedule} \label{sec:timing}
\fref{fig:timeline} illustrates the sequence of operations carried out within one iteration of Algorithm \ref{alg1}, by each of the four main modules detailed in \fref{sec:arch_details}. The horizontal axis shows the progress of time measured in clock cycles. Each row of this diagram corresponds to one of the modules, indicated by its name on the right. The small boxes in each row show the operations of one clock cycle of the corresponding module (if it is active). Each box, contains the antenna $(b)$ and frequency $(w)$ indices of the data being processed in that clock cycle, as well as the output of the module in that cycle (if it exists). Additionally, the diagram shows the timing schedule of the modules relative to each other. For example, at the beginning of each iteration, the MVM1 module begins to compute the first entry of $\bmz_1$, i.e., $[\bmz_1]_1$, after a latency of $L_{\text{UPC}}$ clock cycles. Subsequently, the FTF module begins its operation once it receives a valid output from the MVM1 module, which requires a latency of $L_{\text{UPC}} + L_{\text{MVM1}}$ clock cycles, counted from the beginning of the iteration. It is important to note that the MVM2 module produces valid outputs only during the last $W$ clock cycles of its operation in each iteration, due to its multiply-accumulate approach for computing the matrix-vector product; see \fref{sec:arch_details} for more details.
Once the result $(\kappa \bG)$ of the MVM2 module is ready, they are transferred to the UPC module in order to update the estimate matrix $\bS^{(k-1)}$ of the previous iteration, according to line \ref{Alg1:updateproject} of Algorithm~\ref{alg1}. In total, each algorithm iteration requires $BW + L_{\text{UPC}} + L_{\text{MVM1}} + L_{\text{FTF}} + L_{\text{MVM2}}$ clock cycles as discussed in \fref{sec:VLSIArch}.}

\begin{remark}

\revision{
Typical values of $B$ and $W$ encountered in massive MU-MIMO-OFDM systems, result in relatively high latency and low per-instance throughput, as seen in \fref{tbl:implresults}. This stems from the sequential nature of the proposed 1BOX algorithm, which exhibits stringent data dependencies caused by the repeated time-to-frequency and frequency-to-time transforms. Therefore, it is non-trivial to decompose our algorithm into independent tasks, as opposed to linear data detectors that transform the frequency-selective system into a set of independent frequency-flat systems. Nevertheless, a straightforward solution to increase the throughput would be to deploy multiple parallel instances that operate on different sets of received data.
Besides that, it is also possible to increase the throughput per instance of the proposed data detector, by using highly-parallel FFT architectures and matrix-vector product engines. Further possible improvements are (i) more aggressive pipelining and (ii) interleaved processing of multiple symbols concurrently in a single instance. Combining these techniques with an ASIC implementation in a modern CMOS technology node might be able to achieve Gb/s throughputs. The design of such improved architectures is left for future work.
}

\end{remark}

\subsection{Fixed-Point Parameters} \label{sec:fixedpoint}

In order to maximize hardware efficiency, we deploy fixed-point arithmetic. The word-length of each signal has been optimized based on BER simulations to minimize the loss compared to a floating-point reference model, while maintaining  low area. 
In what follows, we use the notation $[x.y]$ to denote the 2's complement binary format of a fixed-point signal that has $x$ integer bits (including the sign bit) and $y$ fractional bits with a total word length of $x+y$ bits. 
The reported word lengths are for each of the real and imaginary components of complex-valued signals. 
The channel entries $[\bH_w]_b$ are in format $[4.4]$. The entries of the vectors $\bmz_b$ are in format $[5.5]$. The entries of the matrices $\bV$,  $\bG$, and $\bS^{(k)}$ are in format $[4.4]$, $[1.7]$, and $[2.7]$, respectively. 
For the entries of $\mymathb{\alpha}_b$, we use format $[5.4]$. The $\tilde{\omega}$ LUT storing the fixed-point values of the function $\tilde{\omega}(t)$ for $t_{n} < t < t_{p}$, consists of $128$ entries  in format $[3.4]$. The LUT containing the values of $1/\tilde{\sigma}$ consists of $256$ entries in format $[2.6]$. \revision{In  \fref{fig:Architecture}, we show the word length of real and imaginary parts of each complex-valued signal with a number next to the signal connection.}  The BER performance corresponding to the fixed-point hardware implementation is shown in \fref{fig:BER}.

\begin{table*}[tp]
	\centering
	\renewcommand{\arraystretch}{1.1}
	\begin{minipage}[c]{1\textwidth}
		\centering
		\caption{FPGA resource breakdown for the proposed 1BOX data detector on a Xilinx Virtex-7 XC7VX690T FPGA.}
		\label{tbl:Breakdown}
\resizebox{0.98\textwidth}{!}{%
		\begin{tabular}{@{}lccccccccccc@{}}
			\toprule
			System dim.  & \multicolumn{5}{c}{$64\times 4$} && \multicolumn{5}{c}{$128\times 8$} \\
			\cmidrule{2-6}  \cmidrule{8-12}
			{Module}         & H-MEM & UPC & MVM1 & FTF     & MVM2  &&  H-MEM & UPC & MVM1 & FTF & MVM2 \\
			\midrule
			{LUTs}           & 32 (1\%)  & 373 (12\%) & 226 (7\%)   & 2\,263 (72\%) & 231 (8\%)    &&  64 (1\%)  &  712 (18\%) &  482 (12\%) &  2\,263 (58\%) &  431 (11\%) \\
			{FFs}            & 0 (0\%)   & 441 (10\%) & 154 (3\%)   & 3\,615 (82\%) & 232 (5\%)  &&  0 (0\%)   &  849 (16\%) &  346 (6\%) &  3\,616 (70\%) &  410 (8\%) \\
			{DSP48 units}    & 0 (0\%)  & 0 (0\%)  & 16 (30\%)    & 20 (40\%)     & 16 (30\%)   &&  0 (0\%)   &  0 (0\%)  &  32 (38\%) &  20  (24\%)   &  32 (38\%)  \\
			{Block RAMs}     & 16 (72\%) & 0 (0\%)  & 0  (0\%)    & 5  (22\%)  & 1.5 (6\%)       &&  64 (89\%)  &  0 (0\%)  &  0 (0\%)  &  5  (7\%)    &  3  (4\%)  \\
			\bottomrule
		\end{tabular}
}		
	\end{minipage}
\end{table*}

\subsection{Implementation Results and Comparison} \label{sec:ImplementationRes}

In order to demonstrate the efficacy of the proposed architecture for the 1BOX algorithm, we now present implementation results on a Xilinx \mbox{Virtex-7} XC7VX690T FPGA. 
\fref{tbl:implresults} summarizes the FPGA implementation results of the 1BOX detector for two system dimensions $B\times U$, i.e., $64\times4$ and $128\times8$. 
We reiterate that these are, to the best of our knowledge, first hardware implementation results of a 1-bit data detector, which renders a fair comparison with existing designs difficult. 
Nevertheless, we provide a comparison with state-of-the-art massive MIMO data detectors that have been designed for massive MU-MIMO systems with high-resolution ADCs. \revision{We reiterate that this comparison is not entirely fair and its purpose is to show the overhead in hardware efficiency caused by the proposed iterative detection algorithm that supports massive MU-MIMO-OFDM systems with 1-bit ADCs. We emphasize that the reference linear data detectors are not specialized to operate in 1-bit systems and therefore perform poorly in such systems; see \fref{sec:SimRes} for the details.}

The design in \cite{wu2014large} implements an L-MMSE data detector for single-carrier frequency-division multiple access (SC-FDMA)-based massive MU-MIMO systems. This design, which is referred to as ``NS-SCFDMA'' in this paper, uses three iterations of Neumann series to perform an approximate matrix inversion per subcarrier.
The design in \cite{wu16} implements an optimized coordinate descent (OCD) algorithm that approximates a box-constrained detection problem for massive MU-MIMO-OFDM systems---frequency-domain conversion circuitry is, however, excluded. 
The design in \cite{wu17} implements  a parallel Gauss-Seidel (PGS) algorithm that iteratively solves an L-MMSE data detection problem for frequency-flat channels. 
In \fref{tbl:implresults}, we also include two of the most recent, state-of-the-art linear detectors for massive MU-MIMO systems, proposed in \cite{liu20} and \cite{tan20}. The design presented in \cite{liu20}, implements a recursive conjugate-gradient-based L-MMSE detector, called RCG, and the design in \cite{tan20} implements an algorithm based on steepest descent and Barzilai-Borwein algorithms, abbreviated to ESDBB, that solves the L-MMSE detection problem iteratively.
All of the compared designs consider a  system with $B = 128$ BS antennas and $U=8$ UEs.
However, the throughputs reported in the reference designs NS-SCFDMA, OCD, PGS and RCG are for $64$-QAM, while we consider $16$-QAM for the 1BOX data detector. Therefore, in \fref{tbl:implresults}, we scale the reported throughput of these reference designs by a factor of $4/6$ so that the throughput of all designs is with respect to  $16$-QAM. We include the throughput for ESDBB as it was reported in~\cite{tan20}, since the modulation scheme was not specified for that design.
We note that PGS, NS-SCFDMA and RCG designs contain hardware to compute post-equalization signal-to-noise-plus-interference ratio (SINR) and log-likelihood ratios (LLRs), which is not present in our design. 
\fref{tbl:implresults} includes two metrics for comparing the hardware efficiency of different designs: (i) throughput per look-up table (LUT) utilization and (ii) throughput per normalized resource consumption (NRC), which is calculated as $\text{NRC} = \text{LUT} + \text{FF} + 280 \times \text{DSP48}$ \cite{chen16}. The NRC is a more accurate measure for assessing the resource consumption than the LUT count alone as it takes into account the other FPGA resources as well.

As shown in  \fref{tbl:implresults}, the throughput and latency results of the 1BOX data detector fall short of the reference data detectors, which are specialized for BS architectures with high-resolution  ADCs. However, our design achieves similar hardware efficiency in terms of throughput/NRC compared to other designs, and even higher than that of PGS and ESDBB. The RCG implementation achieves the highest throughput and efficiency, but excludes OFDM processing circuitry. 
It is important to discern that the proposed 1BOX implementation directly operates on 1-bit measurements, while the reference designs perform linear data detection based on high-resolution measurements. Consequently, the error-rate performance of the reference designs would, at best, be close to that of ZF-DET shown in \fref{sec:SimRes} for 1-bit massive MU-MIMO systems. Additionally, our 1BOX implementation is designed for OFDM systems and includes DFT processing, while the other designs do not include DFT processing---except for NS-SCFDMA, which includes logic for SC-FDMA processing.  As a result, their hardware efficiency would be lower if OFDM processing circuitry would be included. 
\revision{We note that it is not possible to exclude the submodules corresponding to OFDM processing  in the FTF module from 1BOX detector to enable a meaningful comparison with the reference designs that do not include OFDM processing. This is due to the fact that the operations in the FTF submodule are an integrated part
of each iteration of 1BOX algorithm and they are not separable from the rest of the algorithm. This is in stark contrast to linear detectors for high-resolution OFDM-based systems, that decompose the system into independent frequency-flat subsystems, and a comparison between detectors designed for OFDM-based systems and detectors designed for frequency-flat systems is possible.}

In order to shed more light on the FPGA resource utilization of the proposed 1BOX data detector,  \fref{tbl:Breakdown} shows a breakdown of resources corresponding to the individual modules. We note that the FTF module consumes the largest portion of resources. In turn, roughly 80\% of the FTF resources are consumed by the FFT and IFFT cores that carry out the time-domain to frequency-domain transform and vice versa. This confirms that a direct comparison with other data detectors that exclude such time-to-frequency conversion is not accurate. In addition, \fref{tbl:Breakdown} shows that most of the block RAMs are used in the H-MEM module, which contains the channel matrices for all subcarriers---the reference FPGA designs in \fref{tbl:implresults} for high-resolution ADCs do not provide information on the amount of storage allocated for channel matrices.

%% file: 5-conclusion.tex

\section{Conclusions} \label{sec:conclusion}
We have proposed a quantization-aware data detection algorithm, called 1BOX, for massive MU-MIMO-OFDM systems operating over frequency-selective channels with 1-bit ADCs. To improve the channel estimates for such architectures, we have also proposed a channel estimation algorithm referred to as NGD-CHEST. We have shown using simulations that NGD-CHEST combined with the 1BOX data detector outperform conventional linear channel estimation and data detection methods in terms of error-rate performance. Furthermore, we have developed a reference VLSI architecture and presented corresponding FPGA implementations for 1BOX detector. Our results demonstrate that the proposed design achieves comparable hardware efficiency to data detectors that have been designed for high-resolution systems. \revisionS{The proposed channel estimation and data detection algorithms are particularly useful in all-digital massive MU-MIMO-OFDM systems operating at mmWave or terahertz (THz) frequencies, which require large antenna arrays and high bandwidths. Other use cases for the proposed algorithms and hardware designs could be radar or imaging systems for automotive or robotics applications that operate at high carrier frequencies (e.g., mmWave or THz), in which the RF power consumption is expected to be a bottleneck due to the high bandwidths and the large number of antennas.}

There are numerous avenues for future work. Each instance of the 1BOX data detector achieves throughputs of tens of Mb/s, which is insufficient for next-generation wireless systems operating at millimeter-wave frequencies. \revision{However, there exist several techniques as outlined in \fref{sec:timing}, that can significantly increase the throughput per instance of our architecture. The design of such high-throughput architectures is part of ongoing work.}
\revision{Another important research direction is to explore the overall performance-complexity-cost trade-off in basestations that use low-resolution data converters. In such a study, the effect of reduced data converter resolution on the overall system power consumption and cost should be considered.}
\revision{Finally, the design of efficient algorithms for uplink timing and frequency synchronization as well as parameter estimation  (such as SNR or noise variance) for systems with 1-bit ADCs, is an important open research problem.}